\documentclass[pra,twocolumn,preprintnumbers,amsmath,amssymb,nofootinbib,floatfix]{revtex4}

\usepackage{graphicx, bm, hyperref}

\makeatletter

\def\graphicscale{\twocolumn@sw{0.3}{0.4}}
\def\graphicthreescale{\twocolumn@sw{0.3}{0.4}}

\begin{document}

\title{Scaling of decoherence and energy flow in interacting quantum spin systems}

\author{Davide Rossini and Ettore Vicari} 

\affiliation{Dipartimento di Fisica dell'Universit\`a di Pisa and INFN, 
  Largo Pontecorvo 3, I-56127 Pisa, Italy}

\date{\today}

\begin{abstract}
We address the quantum dynamics of a system composed of a qubit
globally coupled to a many-body system characterized by short-range
interactions.  We employ a dynamic finite-size scaling framework to
investigate the out-of-equilibrium dynamics arising from the sudden
variation (turning on) of the interaction between the qubit and the
many-body system, in particular when the latter is in proximity of a
quantum first-order or continuous phase transition.  Although the
approach is quite general, we consider $d$-dimensional quantum Ising
spin models in the presence of transverse and longitudinal fields, as
paradigmatic quantum many-body systems.  To characterize the
out-of-equilibrium dynamics, we focus on a number of
quantum-information oriented properties of the model.  Namely, we
concentrate on the decoherence features of the qubit, the energy
interchanges among the qubit and the many-body system during the
out-of-equilibrium dynamics, and the work distribution associated with
the quench.  The scaling behaviors predicted by the dynamic
finite-size scaling theory are verified through extensive numerical
computations for the one-dimensional Ising model, which reveal a fast
convergence to the expected asymptotic behavior with increasing the
system size.
 
\end{abstract}

\maketitle

% ========================= BODY =========================

\section{Introduction}
\label{intro}

The recent progress which has been achieved in the control and
manipulation of complex systems at the nano scale has enabled a wealth
of unprecedented possibilities aimed at addressing the unitary quantum
evolution of many-body objects.  These range from the (nearly)
adiabatic dynamics induced by a slow change in time of one of the
control parameters, to the deep out-of-equilibrium dynamics following
an abrupt quench in the system~\cite{Dziarmaga-10, PSSV-11, NH-15}.
In the latter scenario, several fundamental issues have been
investigated, including the onset of thermalization at long times,
quantum transport, and localization phenomena due to the mutual
interplay between disorder and interactions~\cite{Kinoshita-06,
  Hofferberth-07, Trotsky-12, Cheneau-12, Schreiber-15, Kaufman-16,
  Smith-16, Maier-19}.  All of them are eventually devoted to
characterize the highly nonlinear response of the system to the drive,
where nonequilibrium fluctuation relations may play a pivotal
role~\cite{EHM-09, Jarzynski-11, CHT-11, Seifert-12}.
%Prototypical platforms in this respect are ultracold atoms in optical lattices,
%trapped ions, or coupled QED cavities.

Closely related to the scenario we are going to focus in the present
paper, we also mention the raising interest in monitoring the coherent
quantum dynamics of mutually coupled systems, with the purpose to
address energy interchanges or the relative decoherence properties
among the various subsystems~\cite{Zurek-03}.  This kind of study is
relevant both to understand whether quantum mechanics can enhance the
efficiency of energy conversion in complex networks~\cite{Caruso-09,
  Lambert-13}, and to devise novel quantum technologies which are able
to optimize energy storage in subportions of the whole
system~\cite{Binder_2015, Campaioli_2017, Le_2018, Ferraro_18,
  Julia-Farre_2018}.  We shall stress that energy flows are likely to
be influenced by the different quantum phases of the system. Moreover
one would expect an enhanced response in proximity of a quantum phase
transition, which requires special attention~\cite{Sachdev-book}.

The aim of this paper is to shed light on this latter issue.  To this
purpose, we consider the simplest scenario where to frame such
analysis: a single qubit globally coupled to a $d$-dimensional quantum
Ising spin model in a transverse and a longitudinal field.  This spin
model acts as a prototypical quantum many-body system since, when
varying the intensity of the two external fields, it may undergo both
first-order and continuous quantum transitions (FOQTs and CQTs,
respectively).  The composite setup belongs to the class of the so
called central-spin models, where one (or few) qubit can be globally
or locally coupled to the environmental system (see, e.g., 
Refs.~\cite{Zurek-82,CPZ-05,%%
QSLZS-06,RCGMF-07,CFP-07,CP-08,Zurek-09,DQZ-11,NDD-12,SND-16,V-18, %%
Verrucchi-19}).  We put
forward a quantitative scaling theory which generalizes the results of
Ref.~\cite{V-18} focused on the decoherence properties of the qubit,
and meanwhile also carefully addresses the statistics of energy flows
among the qubit and the many-body system.  Specifically, we employ the
finite-size scaling (FSS) framework, which has been shown to be able
to predict the behavior of a system in proximity of either a
CQT~\cite{CPV-14} or a FOQT~\cite{CNPV-14}, as well as in a dynamic
context~\cite{PRV-18b, PRV-18}, providing the asymptotic large-size
scaling in a variety of situations.  A consistent part of this work is
devoted to a numerical validation of the dynamic FSS predictions
through extensive simulations with exact diagonalization techniques,
specialized to the one-dimensional case ($d=1$).

In our setup, the global system is initialized in a state which is a
product of pure states of the qubit $q$ and of the quantum Ising
system $S$.  The qubit $q$ is then suddenly coupled to all the $L^d$
spins of $S$, such that a nontrivial unitary dynamics sets in.  Notice
that we admit the possibility to have an interaction Hamiltonian which
does not commute with the qubit Hamiltonian, so that the decoherent
effect on the qubit is not only a pure dephasing.  We focus on three
quantum-information oriented properties of the model: the decoherence
features of $q$, the statistics of the work distribution associated
with the quench, and the statistics of the energy interchanges among
$q$ and $S$ during the out-of-equilibrium dynamics.  We show that such
properties develop dynamic FSS behaviors when the system $S$ is close
to quantum transitions.  In particular, our numerics for the
one-dimensional Ising systems show that the convergence to the dynamic
FSS behavior is remarkably fast, a fact which encourages a careful
assessment of the role of criticality in near-future experiments of
quantum transport in complex systems with few spins or particles (of
the order of $10$).

The paper is structured as follows.  We start, in Sec.~\ref{gset},
with defining the model and all the relevant quantities that will be
analyzed.  In Sec.~\ref{dfssfra} we summarize the derivation of the
dynamic FSS framework, which is capable to address the dynamics of a
quantum many-body system, either at FOQTs or at CQTs. We also discuss
differences which emerge when considering a disordered phase.  The
dynamic FSS is then specialized to our system of interest in
Sec.~\ref{dfss}, explicitly discussing the qubit decoherence
functions, the work associated with the initial quench, and the
dynamics of the qubit-system energy flow.  Later we enter the details
of the FSS behavior at the CQT point, where extensive numerical
simulations in support of the theory are presented
(Sec.~\ref{numresCQT}), and along the FOQT line, where we also
consider a two-level reduction of the many-body system
(Sec.~\ref{numresFOQT}).  Finally, Sec.~\ref{sec:concl} is devoted to
a summary and perspectives of this work.  In the Appendices we provide
some analytic insight for the special case in which the qubit
Hamiltonian commutes with the qubit-system interaction term, and
discuss more in detail some limitations emerging in the two-level
approximation at the FOQT.

\section{General setting of the problem}
\label{gset}

\subsection{The model}
\label{model}

Let us consider a $d$-dimensional quantum many-body system ($S$) of
size $L^d$, with Hamiltonian
\begin{equation}
H_S(h) = H_c + H_h,\qquad H_h= h \,P \,,
\label{hsdef}
\end{equation}
where $P$ is the spatial integral of local operators, such that
$[H_c,P]\neq 0$, and the parameter $h$ drives a quantum transition
located at $h=0$.
As paradigm example, we shall focus on the quantum Ising model on a
$L^d$ lattice,
\begin{equation}
H_{c} = H_{\rm Is} = 
 - J \, \sum_{\langle {\bm x},{\bm y}\rangle} \sigma^{(3)}_{\bm x}
\sigma^{(3)}_{\bm y} - g\, \sum_{\bm x} \sigma^{(1)}_{\bm x}\,,
\label{hisdef}
\end{equation}
where $\sigma^{(k)}$ are the Pauli matrices, the first sum is over all
bonds connecting nearest-neighbor sites $\langle {\bm x},{\bm
  y}\rangle$, while the other sum is over all sites.  Hereafter we
assume $\hslash=1$, $J=1$, the lattice spacing $a=1$, and $g>0$.  At
$g=g_c$ (in one dimension, $g_c=1$), the model undergoes a CQT
belonging to the $(d+1)$-dimensional Ising universality
class~\cite{Sachdev-book, ZJ-book, PV-02}, separating a disordered
phase ($g>g_c$) from an ordered one ($g<g_c$).  The presence of a
homogeneous longitudinal external field is taken into account by
adding the term
\begin{equation}
  h \,P =  - h \sum_{\bm x} \sigma^{(3)}_{\bm x}
  \label{hvisdef}
\end{equation}
to the Hamiltonian (\ref{hisdef}).  The field $h$ drives FOQTs along
the $h=0$ line for any $g<g_c$. At the continuous transition, $g=g_c$,
such term is one of the relevant perturbations driving the critical
behavior, the other one being the transverse field term,
$(g-g_c)\sum_{\bm x} \sigma^{(1)}_{\bm x}$.

In addition, let us consider a qubit ($q$) whose two-level Hamiltonian
can be generally written as
\begin{equation}
H_q = \sum_{a=\pm} \epsilon_a | a \rangle \langle a | = \alpha\, I_2 + 
\tfrac12 \Sigma^{(3)} ,
\label{Hqdef}
\end{equation}
where $I_2$ is the $2 \times 2$ identity matrix, and the Pauli operator
$\Sigma^{(3)}$ is associated with the two states $|\pm\rangle$ of the
qubit, so that $\Sigma^{(3)} | \pm \rangle = \pm | \pm \rangle$.
Therefore
\begin{equation}
\epsilon_\pm = \alpha \pm \tfrac12 \delta, \qquad \delta = \epsilon_+ -
\epsilon_-\,.
\label{epspm}
\end{equation}
The qubit is globally and homogeneously coupled to the many-body
system $S$, through the Hamiltonian term
\begin{equation}
H_{qS} = \left( u \,\Sigma^{(3)} + v \, \Sigma^{(1)} \right) P\,,
\label{hqdef}
\end{equation}
where $P$ is the operator appearing in Eq.~(\ref{hsdef}), thus
Eq.~(\ref{hvisdef}). Putting all the terms together, we obtain the
global Hamiltonian
\begin{equation}
H = H_S + H_q + H_{qS}.
\label{hlamu}
\end{equation}

We are interested in the quantum evolution of the global system
starting from the initial, $t=0$, condition
\begin{equation}
|\Psi_0 \rangle = | q_0 \rangle \otimes | 0_h \rangle\,,
\label{psit0}
\end{equation}
where $|q_0\rangle$ is a generic pure state of the qubit,
\begin{equation}
|q_0 \rangle = c_+ |+\rangle  + c_- |-\rangle,\qquad 
|c_+|^2 + |c_-|^2 = 1\,,
\label{iqstate}
\end{equation} 
and $| 0_h \rangle$ is the ground state of the system with Hamiltonian
$H_S(h)$.  The global wave function describing the quantum evolution
for $t>0$ must be solution of the Schr\"odinger equation
\begin{equation}
i {\partial 
\over \partial t} |\Psi(t)\rangle = H
|\Psi(t)\rangle\,, \qquad |\Psi(t=0)\rangle=|\Psi_0\rangle\,.
\label{sceq}
\end{equation}
In particular, we consider a dynamic protocol arising from a sudden
switching of the interaction $H_{qS}$ between the qubit and the
many-body system, at time $t=0$, i.e., by quenching one or both of the
control parameters $u$ and $v$ in Eq.~\eqref{hqdef} from zero to some
finite value. 

The above setting can be straightforwardly extended to $N$-level
systems coupled to an environmental system $S$, and also to the case
the initial qubit state is mixed, thus described by a nontrivial
density matrix. Most features addressed in the rest of the paper, in
particular the dynamic scaling properties, can be straightforwardly
extended as well.

\subsection{Qubit decoherence, work, and qubit-system energy exchanges}
\label{definitions} 

\subsubsection{Coherence of the qubit}
\label{cohpro}

An interesting issue arising from the dynamics of the problem outlined
above concerns the coherence properties of the qubit during the global
quantum evolution.  Starting from a pure state, the interaction with
the many-body system may give rise to a loss of coherence of the
qubit, depending on the properties of its density matrix
\begin{equation}
\rho_q(t) = {\rm Tr}_S [\rho(t)]\,,\quad 
\label{rhoq}
\end{equation}
where ${\rm Tr}_S[\:\cdot\:]$ denotes the trace over the
$S$-degrees-of-freedom of the global (pure) quantum state
\begin{equation}
\rho(t)=|\Psi(t)\rangle \langle\Psi(t)|\,, 
\label{rhot}
\end{equation}
with $|\Psi(t)\rangle$
given by the solution of Eq.~\eqref{sceq}.  Of course, ${\rm Tr}[
  \rho_q(t)] = 1$.

A good candidate to quantify the coherence properties of the qubit
during its quantum evolution is provided by the so-called {\em
  purity}, that is, the trace of its square density matrix $\rho_q$,
\begin{equation}
{\rm Tr}\, \big[ \rho_q(t)^2 \big] \equiv 1 - D(t) \,,
\label{ddef}
\end{equation}
where we have introduced the decoherence function $D$, such that $0\le
D \le 1$.  This function measures the quantum decoherence, quantifying
the departure from a pure state.  Indeed $D=0$ implies that the qubit
is in a pure state, thus $D(t=0)=0$.  The other extreme value $D=1$
indicates that the qubit is totally unpolarized.

\subsubsection{Quantum work associated with the initial quench}
\label{quwo}

The initial quench, arising from turning on the interaction between
the qubit $q$ and the many-body system $S$, can be also characterized
via the quantum work $W$ done on the global system~\cite{CHT-11,
  GPGS-18}.  The work performed by quenching the control parameters
$u$ and $v$ does not generally have a definite value, while it can be
defined as the difference of two projective energy
measurements~\cite{CHT-11}.  The first one at $t=0$ projects onto the
eigenstates $|m_{\rm i}\rangle$ of the initial Hamiltonian $H_{\rm i}$
with a probability $p_{{\rm i},m}$ given by the initial density
matrix.  The second energy measurement projects onto the eigenstates
$|n_{\rm f}\rangle$ of the post-quench Hamiltonian $H_{\rm f}$.  Since
the energy is conserved after the quench, the latter measurement can be performed
at any time $t$ during the evolution, ruled by the unitary operator
$U(t,0)=e^{-i H_{\rm f} t}$, without changing the distribution, in
particular for $t\to 0^+$.
The work probability distribution can be written
as~\cite{CHT-11, TH-16, TLH-07}:
\begin{equation}
  P(W) = \sum_{n,m} \delta \big[ W-(E_{{\rm f},n}-E_{{\rm i},m}) \big] \,
  \big| \langle n_{\rm f} | m_{\rm i} \rangle \big|^2 \,
  p_{{\rm i},m}\,.
  \label{pwdefft}
\end{equation}
The average work and its higher moments are given by
\begin{equation}
\langle W^k \rangle  = \int dW\, W^k\, P(W) \,.
\label{scalwk}
\end{equation}

As one can easily check, the average quantum work $\langle W \rangle$
can be computed by taking the difference
\begin{equation}
\langle W \rangle = \langle \Psi(t)| H |\Psi(t)\rangle -
\langle \Psi_0| H_q+H_S |\Psi_0\rangle \,,
\label{workdef}
\end{equation}
where
\begin{equation}
\langle \Psi_0| H_q + H_S |\Psi_0\rangle
= \langle q| H_q | q \rangle +
\langle 0_h | H_S | 0_h \rangle \equiv E_{q0} + E_{S0}\,.
\end{equation}
Since we are interested in a sudden quench at $t=0$, we can obtain the
average work from the difference of the expectation values of $H_{\rm
  f}=H$ and $H_{\rm i}=H_q+H_S$ on the initial state, obtaining
\begin{equation}
\langle W \rangle  = \langle \Psi_0| H_{qS}| \Psi_0\rangle\,.
\label{eqs0}
\end{equation}
An analogous expression can be derived for the average of the square
work, obtaining
\begin{equation}
\langle W^2 \rangle = \langle \Psi_0| H_{qS}^2 | \Psi_0\rangle\,.
\label{sqwork}
\end{equation}
Note that the above relatively simple equations for the first two moments
of the work distribution do not extend to higher moments, i.e.
$\langle W^k \rangle \neq \langle \Psi_0| H_{qS}^k | \Psi_0\rangle$
for $k>2$.  Their expressions are more complicated, requiring the
computation of the whole spectrum, due to the fact that $H_{qS}$ does
not commute with the other Hamiltonian terms.

\subsubsection{Energy-difference distributions}
\label{endisr}

In order to study the qubit-system energy exchanges, we may consider
the energy-difference distribution of the system $S$ along the quantum
evolution, associated with two energy measurements of $S$, at $t=0$
and at a generic time $t$.  We write it as
\begin{equation}
  P_S(U, t) = \!\! \sum_{n,a,b} \! \delta \big[ U-(E_{Sn}-E_{S0})\big]
  \, \big| \langle b,n | e^{-iH t} | a,0 \rangle \big|^2 p_a,
  \label{pdedef}
\end{equation}
where $|b,n\rangle\equiv |b\rangle\otimes|n\rangle$, $|n \rangle$ and
$E_{Sn}$ indicate eigenstates and eigenvalues of the Hamiltonian $H_S$
(we assume a discrete spectrum, as is generally appropriate for
finite-size systems), $|a\rangle, |b\rangle$ indicate the eigenstates
$|\pm\rangle$ of the qubit Hamiltonian, and $p_\pm = |c_\pm|^2$ are
the probabilities of the initial qubit state at $t=0$.  One can check
that
\begin{equation}
  \langle U \rangle \equiv  \int dy \,y\,P(y,t) = E_S(t) - E_{S0} \, .
  \label{deltastpet}
\end{equation}
More general initial distributions may be also considered, for example
associated with a bath at temperature $T$, replacing the initial
density matrix $\rho_0 = | q,0\rangle\langle q,0|$ with the
corresponding density matrices.

Of course one may also define an analogous energy-difference
distribution associated with the qubit, obtained by a two-measurement
procedure on it:
\begin{equation}
  P_q(U, t) = \! \sum_{n,a,b} \delta \big[ U - (E_{qb}-E_{qa})\big]
  \, \big| \langle b, n | e^{-iH t} | a, 0 \rangle \big|^2 p_a\,. 
  \label{pdedefq}
\end{equation}

%Note that Eq.~(\ref{pdedef}) can be also written as
%\begin{eqnarray}
%&& P({\cal E}, t) =\sum_{na} \delta \big[ {\cal
%      E}-(E_{sn}-E_{s0})\big] \, {\rm Tr}\,[\rho_{na}\,\rho(t)], \quad
%  \label{pdedef2}\\
% &&\rho_{na} = |n\,a\rangle\langle n\,a|\,,\quad
% \rho(t) = e^{-iHt}\,|0\ q\rangle \langle 0\, q|\,e^{iHt}\,.
% \nonumber
%\end{eqnarray}

\section{The finite-size scaling framework}
\label{dfssfra}

In this section we summarize the main features of the dynamic FSS
framework that we will exploit to analyze the out-of-equilibrium
quantum dynamics of the coupled qubit and many-body system, and in
particular the quantities introduced in Sec.~\ref{definitions}.  The
dynamic FSS framework has been recently developed to deal with dynamic
behaviors of finite-size systems at quantum transitions~\cite{PRV-18b,
  V-18, NRV-18}, extending equilibrium FSS frameworks to study CQTs
and FOQTs~\cite{CPV-14, CNPV-15, CPV-15, PRV-18c, RV-18}.  This
framework allows one to study the interplay among the Hamiltonian
parameters, the finite linear size $L$, and the finite temperature
$T$, assuming that $T$ is sufficiently small, and that the Hamiltonian
parameters keep the system $S$ close to the transition point.

\subsection{Scaling variables at quantum transitions}
\label{scalvar}

The scaling hypothesis for the many-body system described by the
Hamiltonian $H_S$ is based on the existence of a nontrivial
large-volume limit, keeping the appropriate scaling variables fixed.
At both CQTs and FOQTs, the FSS variable related to a relevant
perturbation as $H_h= h P$, cf. Eq.~(\ref{hsdef}), can be generally
written as the ratio
\begin{equation}
\kappa_h = E_h(L) / \Delta(L),
\label{kappah}
\end{equation}
between the energy variation associated with the $H_h$ term (we assume
$E_h= 0$ at the transition point $h=0$) and the energy difference of the
lowest-energy states, $\Delta(L)\equiv E_1 - E_0$, at the transition
point $h=0$.  Nonzero temperatures are taken into account by adding a
further scaling variable
\begin{equation}
\tau = T/\Delta(L)\,.
\label{taudef}
\end{equation}
More generally, any relevant low-energy energy scale ${\cal E}$ is
expected to behave as ${\cal E}\sim \Delta(L)$ in the FSS limit.
Dynamic behaviors, exhibiting nontrivial time dependencies,
also require a scaling variable associated with the time variable,
which is generally given by
\begin{equation}
\theta = \Delta(L)\, t\,.
\label{thetadef}
\end{equation}
The equilibrium and dynamic FSS limits are defined as the large-size
limit, keeping the above scaling variables fixed.

The outlined framework provides a unified picture of the FSS behaviors
at quantum transitions, holding at both CQTs and FOQTs.  Within the
dynamic FSS framework, their differences are essentially related to
the functional dependence of the above scaling variables on the size.
Power laws generally arise at CQTs~\cite{CPV-14}, while exponential
laws emerge at FOQTs~\cite{CNPV-15, PRV-18c}, in particular when
boundary conditions do not favor any particular phase.

\subsection{First-order quantum transitions}
\label{scalvarfoqt}

As shown by earlier works~\cite{CNPV-14, CNPV-15, CPV-15, PRV-18c},
the FSS behavior of isolated many-body systems at FOQTs turns out to
be much depending on the type of boundary conditions, whether they
favor one of the phases or they are neutral, giving rise to FSS
characterized by exponential or power-law behaviors.  To simplify our
presentation, in the following, the system $S$ will be taken as a
quantum Ising model with boundary conditions that do not favor any of
the two magnetized phases, such as periodic and open boundary
conditions, which generally lead to exponential FSS laws.

The FOQT line for $g<g_c$ is related to the level crossing of the two
lowest states $| \! \uparrow \rangle$ and $| \! \downarrow \rangle$
for $h=0$, such that $\langle \uparrow \! |\sigma_{\bm x}^{(3)} | \!
\uparrow \rangle = m_0$ and $\langle \downarrow \! | \sigma_{\bm
  x}^{(3)} \! | \downarrow \rangle = -m_0$ (independently of ${\bm
  x}$), with $m_0>0$.  The degeneracy of these states is lifted by the
longitudinal field $h$.  Therefore, $h = 0$ is a FOQT point, where the
longitudinal magnetization $M = L^{-d} \sum_{\bm x} M_{\bm x}$, with
$M_{\bm x}\equiv \langle \sigma_{\bm x}^{(3)} \rangle$, becomes
discontinuous in the infinite-volume limit.  The transition separates
two different phases characterized by opposite values of the
magnetization $m_0$, i.e.
\begin{equation}
  \lim_{h \to 0^\pm} \lim_{L\to\infty} M = \pm m_0\,.
  \label{m0def}
\end{equation}
For one-dimensional systems~\cite{Pfeuty-70}, $m_0 = (1 - g^2)^{1/8}$.

In a finite system of size $L$, the two lowest states are
superpositions of two magnetized states $| + \rangle$ and $| -
\rangle$ such that $\langle \pm | \sigma_{\bm x}^{(3)} | \pm \rangle =
\pm \,m_0$ for all sites ${\bm x}$.  Due to tunneling effects, the
energy gap $\Delta$ at $h=0$ vanishes exponentially as $L$
increases~\cite{PF-83, CNPV-14},
\begin{equation}
  \Delta(L) \sim e^{-c L^d}\,,
  \label{del}
\end{equation}
apart from powers of $L$.  In particular, for the one-dimensional
Ising system (\ref{hisdef}) at $g<1$, it is exponentially suppressed
as~\cite{Pfeuty-70, CJ-87}
\begin{equation}
  \Delta(L) = 2 \, (1-g^2) g^L \, [1+ O(g^{2L})]
  \label{deltaobc}
\end{equation}
for open boundary conditions, and
\begin{equation}
  \Delta(L) \approx 2 \, \sqrt{(1-g^2)/(\pi L)} \, g^L
  \label{deltapbc}
\end{equation}
for periodic boundary conditions.  The differences $E_i-E_0$ for the
higher excited states $(i>1$) are finite for $L\to \infty$.

Quantum Ising systems along the FOQT line develop FSS
behaviors~\cite{PRV-18, PRV-18b, V-18, NRV-18}, driven by the
longitudinal field $h$.  Using Eq.~\eqref{kappah}, the corresponding
scaling variable can be written as~\cite{CNPV-15}
\begin{equation}
  \kappa_h = {2 m_0 h \,L^d \over \Delta(L)} \,,
  \label{scavarfoqt}
\end{equation}
where $2m_0 h L^d$ approximately quantifies the energy associated with
the corresponding longitudinal-field perturbation $H_h$.  For example,
in the equilibrium FSS limit the magnetization is expected to behave
as~\cite{CNPV-15} $M(h,L) = m_0 \, {\cal M}(\kappa_h)$, where ${\cal
  M}$ is a FSS function.

Note that the FOQT scenario based on the avoided crossing of two
levels is not realized for any boundary condition~\cite{CNPV-14}: in
some cases the energy difference $\Delta(L)$ of the lowest levels may
even display a power-law dependence on $L$.  However, the scaling
variable $\kappa_h$ obtained using the corresponding $\Delta(L)$ turns
out to be appropriate, as well~\cite{CNPV-14}.

\subsection{Continuous quantum transitions}
\label{scalvarcqt}

FSS theories have been originally developed at continuous transitions,
see e.g. Refs.~\cite{Barber-83, Privman-90, CPV-14} and references
therein.  The CQT of the Ising model, cf. Eqs.~(\ref{hisdef})
and~\eqref{hvisdef}, is characterized by two relevant parameters,
$r\equiv g-g_c$ and $h$ (such that they vanish at the critical point),
with renormalization-group dimension $y_r$ and $y_h$, respectively.
The relevant FSS variables are
\begin{equation}
  \kappa_r = L^{y_r} r\,,\qquad \kappa_h= L^{y_h} h\,.
  \label{karh}
\end{equation}
The FSS limit is obtained by taking $L\to\infty$,
keeping $\kappa_r$ and $\kappa_h$ fixed.

Note that the expression for $\kappa_h$ in Eq.~\eqref{karh} can be
obtained using the more general definition (\ref{kappah}).  Indeed, at
CQTs the energy variation arising from the perturbation $H_h = h
\sum_{\bm x} P_{\bm x}$, with $P_{\bm x} = - \sigma^{(3)}_{\bm x}$, is
given by
\begin{equation}
  E_h(L) \sim h L^{d - y_p} \,,
  \label{ehl}
\end{equation}
where $y_p$ is the renormalization-group critical dimension of the
local operators $P_{\bm x}$ at the fixed point describing the quantum
critical behavior.  Moreover, we have
\begin{equation}
  \Delta(L) \sim L^{-z} \,,
  \label{deltaz}
\end{equation}
where $z$ is the universal dynamic exponent.  Then, using the scaling
relation among critical exponents~\cite{CPV-14,Sachdev-book}
\begin{equation}
  y_h + y_p = d + z \,,
  \label{scalrel}
\end{equation}
where $y_h$ is the RG dimension of the perturbation $h$, we end up
with the expression of $\kappa_h$ reported in Eq.~(\ref{karh}).  An
analogous derivation can be obtained for $\kappa_r$.

The equilibrium critical exponents $y_r$ and $y_h$ of the quantum
Ising model are those of the $(d+1)$-dimensional Ising universality
class~\cite{Sachdev-book, ZJ-book, PV-02}. Therefore, for
one-dimensional systems they are $y_r=1/\nu=1$ and $y_h =
(d+3-\eta)/2= (4-\eta)/2$ with $\eta=1/4$.  For two-dimensional models
the critical exponents are not known exactly, but there are very
accurate estimates (see, e.g., Refs.~\cite{GZ-98, CPRV-02,
  Hasenbusch-10, KPSV-16, KP-17}); in particular~\cite{KPSV-16}
$y_r=1/\nu$ with $\nu=0.629971(4)$ and $y_h = (5-\eta)/2$ with
$\eta=0.036298(2)$.  For three-dimensional systems they assume
mean-field values, $y_r=2$ and $y_h=3$, apart from logarithms.  The
temperature $T$ gives rise to a relevant perturbation at CQTs,
associated with the scaling variable $\tau=L^z T$, where $z=1$ (for
any spatial dimension) is the dynamic exponent characterizing the
behavior of the energy differences of the lowest-energy states and, in
particular, the gap $\Delta\sim L^{-z}$.

A generic observable $O$ in the FSS limit behaves as
\begin{eqnarray}
 O(r,h, L) \approx L^{-y_o} \, {\cal O}(\kappa_r, \kappa_h)\,,
  \label{cqtequi}
\end{eqnarray}
where the exponent $y_o$ is the renormalization-group dimension
associated with $O$, and ${\cal O}$ is a universal equilibrium FSS
function.  The approach to such an asymptotic behavior is
characterized by power-law corrections, typically controlled by
irrelevant perturbations at the corresponding fixed
point~\cite{CPV-14}.  The equilibrium FSS at quantum transitions has
been also extended to quantum-information concepts~\cite{GPSZ-06,
  AFOV-08, Gu-10, BAB-17, DS-18}, such as the ground-state fidelity
and its susceptibility, which measure the change of the ground state
when varying the Hamiltonian parameters around a quantum
transition~\cite{RV-18}.

\subsection{The disordered phase}
\label{disordered}

One may compare the above scaling behaviors with those expected when
the system $S$ is not close to a phase transition, for example for
$g>g_c$ in the case of quantum Ising models. In this region the
system is in the disordered phase, where the length scale $\xi$ of the
correlations is finite. In particular, close to the transition point
$g_c$, it behaves as $\xi\sim (g-g_c)^{-\nu}$.  Thus the ratio $L/\xi$
diverges in the large-$L$ limit.  The many-body system appears as
effectively composed of $(L/\xi)^d$ uncorrelated subsystems. The gap
$\Delta(L)$ remains finite with increasing $L$. Close to the CQT,
i.e. for $g\gtrsim g_c$, it behaves as $\Delta\sim \xi^{-z}$.

\section{Dynamic FSS ansatz for the qubit-system setup}
\label{dfss}

The dynamic processes arising from the instantaneous turning on of the
interaction term $H_{qS}$ can be described within a dynamic FSS
framework, extending the framework outlined in Sec.~\ref{dfssfra}, to
take into account the interaction of the many-body system $S$ with the
qubit $q$. Beside the scaling variable $\kappa_h$,
cf. Eq.~(\ref{kappah}), we also need to consider scaling variables
associated with the other parameters of the global Hamiltonian $H$,
i.e., $u$, $v$ and $\delta$.

Since both the $u$- and $v$-term are coupled to the operator $P$
contained in the $h$-term of the global Hamiltonian~\eqref{hsdef}, we
expect the corresponding scaling variables to scale analogously as
$\kappa_h$. They can thus be obtained by replacing $h$ with $u$ (or
$v$) in the definition of $\kappa_h$.  Given that $\kappa_h$ is linear
in $h$, one has
\begin{equation}
\kappa_u = u \,\kappa_h/h\,, \qquad
\kappa_v = v \,\kappa_h/h\,. 
\label{kappauv}
\end{equation}
We must also associate a FSS variable with the energy difference
$\delta$ of the eigenstates of the qubit Hamiltonian $H_q$,
\begin{equation}
  \varepsilon_\delta = {\delta/\Delta(L)} \,,
  \label{varepsilon}
\end{equation}
since $\delta$ is a further energy scale of the problem.
Finally, the scaling variable $\theta=\Delta(L)\,t$ is associated with
the time variable.
%The dynamic FSS limit is defined as the large-$L$ limit
%keeping the above scaling variables fixed.

\subsubsection{Qubit decoherence functions}
\label{decoqubit}

Let us first address the issue of decoherence properties of the qubit
along the global quantum evolution arising from the interaction with
the system $S$.
The dynamic scaling behavior of the decoherence function $D(t)$,
cf. Eq.~(\ref{ddef}), as a function of time, size, and Hamiltonian
parameters of the system $S$, has been discussed in Ref.~\cite{V-18}
in the simplest case where $\delta = 0$ (i.e., the qubit Hamiltonian
is trivial).  As detailed in App.~\ref{commcase}, each time
$[H_q,H_{qS}]=0$ [as is the case for $v=0$ in Eq.~(\ref{hqdef})],
$D(t)$ does not depend on the qubit spectrum.  We thus expect the same
dynamic FSS behavior reported in Ref.~\cite{V-18},
\begin{equation}
  D(u,h,L,t) = {\cal D}(\kappa_u,\kappa_h,\theta) \,,
  \label{calpscal}
\end{equation}
which is independent of $\delta$, and therefore of $\varepsilon_\delta$.
For small values of the coupling $u$, we have that
\begin{equation}
  D(u,h,L,t) = \tfrac12 u^2 Q(h,L,t) + O(u^4) \,,
  \label{calqdef}
\end{equation}
where the growth-rate function $Q$ measures the sensitivity of the
qubit coherence properties to the coupling $u$.  Its scaling behavior
can be derived by matching that of ${\cal D}$ in Eq.~(\ref{calpscal}),
obtaining~\cite{V-18}
\begin{equation}
  Q(h,L,t) \approx \left( {\partial \kappa_u\over\partial u}\right)^2 
  {\cal Q}(\kappa_h,\theta) \,.
  \label{claQscal}
\end{equation}

We now focus on the more general case $[H_q,H_{qS}]\neq 0$,
restricting to the case $u=0$ for simplicity, without loss of
generality.  The most natural working hypothesis is that analogous
scaling behaviors develop, with a further dependence on the scaling
variable $\varepsilon_\delta$ in Eq.~\eqref{varepsilon}, and replacing
$\kappa_u$ with $\kappa_v$.  One is then led to put forward the
following FSS behavior:
\begin{equation}
  D(\delta,v,h,L,t) = {\cal D}(\varepsilon_\delta,\kappa_v,\kappa_h,\theta)\,.
  \label{calpscalnc}
\end{equation}
Assuming again analyticity at $v=0$ and since $D\geq 0$, one expects
an expansion analogous to Eq.~\eqref{calqdef},
\begin{equation}
  D(\delta,v,h,L,t) = \tfrac12 v^2 Q(\delta,h,L,t) + O(v^4),
  \label{calqdefv}
\end{equation}
which can be matched to the scaling behavior in
Eq.~\eqref{calpscalnc}, to obtain
\begin{eqnarray}
Q(\delta,h,L,t) \approx \left( {\partial \kappa_v\over\partial
  v}\right)^2 {\cal Q}(\varepsilon_\delta,
\kappa_h,\theta)\,.
\label{claQscalnc}
\end{eqnarray}

Notice that the scaling relations for the growth-rate function $Q$
imply a power law for CQTs, i.e.
\begin{equation}
Q(h,L,t) \approx L^{2y_h} {\cal Q}(\varepsilon_\delta,\kappa_h,\theta),
\label{claQscalcqt}
\end{equation}
while exponential laws arise at FOQTs (when considering neutral boundary
conditions), such as
\begin{equation}
Q(h,L,t) \approx  {L^{2d}\over \Delta(L)^{2}}
{\cal Q}(\varepsilon_\delta,\kappa_h,\theta),
\label{claQscalfoqt}
\end{equation}
thus increasing as $\sim\exp(b L^d)$.

One may compare the above scaling behaviors with those expected when
the system $S$ is not close to a phase transition, for example for
$g>g_c$ in the case of the quantum Ising models.  Keeping into account
that $Q$ is equivalent to a generalized susceptibility, the arguments
reported in Sec.~\ref{disordered} lead us to the expectation that the
growth-rate function should increase as the volume of the system $S$,
i.e.
\begin{equation}
  Q \sim L^{d}\,.
  \label{claQscalnocri}
\end{equation}
The above scaling relations~\eqref{claQscalcqt}
and~\eqref{claQscalfoqt} demonstrate that the rate of the qubit
decoherence gets enhanced when the system $S$ experiences a quantum
transition.  For example, we may compare the $L^d$ behavior expected
in the disordered phase with the significantly faster increase
$L^{2y_h}$ at $g=g_c$ where the system is critical,
cf. Eq.~(\ref{claQscalcqt}), due to the fact that ($z=1$)
\begin{equation}
2y_h=d+z+2-\eta=d+3-\eta>d\,.
\label{2yhltd}
\end{equation}

We finally note that the scaling behavior in the critical region,
$g-g_c\ll 1$, where $\xi\approx (g-g_c)^{-\nu}\gg 1$, can be inferred
by standard scaling arguments.  It essentially amounts to replace $L$
with $\xi$ in the dynamic FSS equations.

\subsubsection{Quantum work associated with the initial quench}
\label{wosca}

We now discuss the scaling behavior of the quantum work associated
with the initial quench of the interaction term $H_{qS}$.
We explicitly address the first and the second moment of
the work probability distribution defined in Eq.~\eqref{pwdefft},
but similar arguments can be put forward for the higher moments,
as defined in Eq~\eqref{scalwk}.

To begin with, we report the scaling ansatzes of the average work and
average square work which are expected in the simplest case
$[H_q,H_{qS}]=0$ (for $v=0$):
\begin{subequations}
  \begin{eqnarray}
    \langle W\rangle(\delta,u,h,L) &\approx&\Delta(L)\, {\cal
      W}_1(\kappa_u,\kappa_h)\,,
    \label{scaloverw}\\
    \langle W^2\rangle(\delta,u,h,L) &\approx& \Delta(L)^2 \,{\cal W}_2(\kappa_u,\kappa_h)\,.
    \label{scaloverw2}
  \end{eqnarray}
\end{subequations}
These are independent of the spectrum of the qubit, and in particular
of $\delta$ (analogous expressions hold for higher powers of the
work).
In the case of CQTs, this ansatz is supported by equilibrium
FSS arguments, exploiting Eq.~(\ref{auwork}). Indeed, since
\begin{equation}
\langle 0_h | P | 0_h \rangle \approx L^{d-y_p} f_P(\kappa_h)\,,
\label{scap}
\end{equation}
it is easy to see that simple calculations lead to
Eq.~(\ref{scaloverw}) with ${\cal W}_1(\kappa_u,\kappa_h) \propto
\kappa_u f_P(\kappa_h)$.

In the more general case $[H_q,H_{qS}]\neq 0$, we expect that the
scaling variable $\varepsilon_\delta$ associated with the gap $\delta$
of the qubit Hamiltonian should also enter the dynamic FSS behavior.
However, for the average and the average square work
[cf. Eqs.~\eqref{eqs0} and~\eqref{sqwork}, respectively], this is not
the case:
\begin{subequations}
\begin{eqnarray}
  \langle W \rangle(\delta,u,v,h,L) &\approx& \Delta(L) \, {\cal
    W}_1(\kappa_u,\kappa_v,\kappa_h)\,,
  \label{scaloverwnc}\\
  \langle W^2 \rangle(\delta,u,v,h,L) &\approx& \Delta(L)^2 \, {\cal
    W}_2(\kappa_u,\kappa_v,\kappa_h)\,.\qquad
  \label{scaloverwnc2}
\end{eqnarray}
\label{scaloverw_bis}%
\end{subequations}
Higher moments are expected to generally depend on
$\varepsilon_\delta$ too.

\subsubsection{Time dependence of the energy exchanges}
\label{enesflusca}

Coming to the energy distribution defined in Eq.~(\ref{pdedef}), let
us again begin with the simplest case $v=0$.  We expect the scaling
behavior
\begin{equation}
  P_S(U,u,h,L,t) \approx \Delta(L)^{-1} \,{\cal
    P}(\upsilon,\kappa_u,\kappa_h,\theta)\,,
\label{pescalc}
\end{equation}
where
\begin{equation}
  \upsilon = {U/\Delta(L)}\,.
  \label{varecaledef}
\end{equation}
Thus the average of $U$ and its fluctuations
$\langle U^2 \rangle_c = \langle U^2 \rangle - \langle U \rangle^2$
should scale respectively as
\begin{subequations}
\begin{eqnarray}
  \langle U\rangle(u,h,L,t) & \approx & \Delta(L) \: {\cal U}_1(\kappa_u,\kappa_h,\theta) \, ,
  \label{escalc} \\
  \langle U^2 \rangle_c(u,h,L,t) & \approx & \Delta(L)^2 \: {\cal U}_2(\kappa_u,\kappa_h,\theta)\, . 
  \label{escalc2}
\end{eqnarray}
\label{escal}%
\end{subequations}

In the most general case $v\neq 0$, one should add a further
dependence on the scaling variable $\varepsilon_\delta$ associated
with the gap of the qubit Hamiltonian. Therefore we obtain
\begin{equation}
  P_S(U,\delta,u,v,h,L,t) \approx \Delta(L)^{-1} {\cal P}(\upsilon,
  \varepsilon_\delta,\kappa_u,\kappa_v,\kappa_h,\theta) \,.
  \label{pescalcgen}
\end{equation}

\section{Dynamic behavior at the CQT}
\label{numresCQT}

We have seen that the dynamic FSS theory specialized to the
qubit-system setup predicts a non trivial scaling limit for the
different properties of the model. Here we focus on the behavior at
the CQT of the one-dimensional Ising model, and present the results of
numerical exact diagonalization simulations for the dynamics of a
qubit, after it is suddenly and homogeneously coupled to an Ising ring
with Hamiltonian~\eqref{hisdef}.  We shall remind the reader that, as
detailed in App.~\ref{commcase}, for $v = 0$ the qubit exhibits pure
dephasing in time, while the system evolves in two independent
branches with the same Hamiltonian form and different
fields~\cite{QSLZS-06, RCGMF-07} [cf. Eq.~\eqref{psitcom}].  A dynamic
FSS theory for the decoherence functions of the qubit coupled to a
many-body system at a quantum transition has been already addressed in
the literature, in this specific case and for $\delta=0$~\cite{V-18}.

Here let us discuss the general case in which the Hamiltonian
coupling term $H_{qS}$ does not commute with the qubit Hamiltonian
$H_q$, i.e. $v \neq 0$ in Eq.~\eqref{hqdef}.  We simulated setups
where the Ising ring is constituted of up to $L=24$ sites for static
calculations, while we limited ourselves to $L=16$ sites for the
dynamics of such systems.  As we shall see below, the dynamic FSS
scaling behavior turns out to emerge quite neatly already for these
moderate lengths, thus making unnecessary, in practice, any further
extensive check at larger size.  A full exact diagonalization approach
has been used for systems with $L\leq 12$, while a Lanczos
diagonalization followed by a fourth-order Runge-Kutta integration of
the unitary-evolution operator was employed for larger sizes ($13 \leq
L \leq 16$).  We carefully checked that, for most of the simulations,
a time step $dt = 10^{-3}$ is sufficient to reach a high degree of
convergence.  As we shall explain below, for the calculation of
second-order temporal fluctuations of the statistics of energy
exchanges, at the largest considered size, a smaller time step $dt =
10^{-4}$ turns out to be required.

%%%%%%%%%%%%%%%%%%%%%%%%%%%%%%%%%%%%%%%%%%%%%%%%%%%%%%%%%%%%%%%%%%%
\begin{figure}[!t]
  \includegraphics[width=0.95\columnwidth]{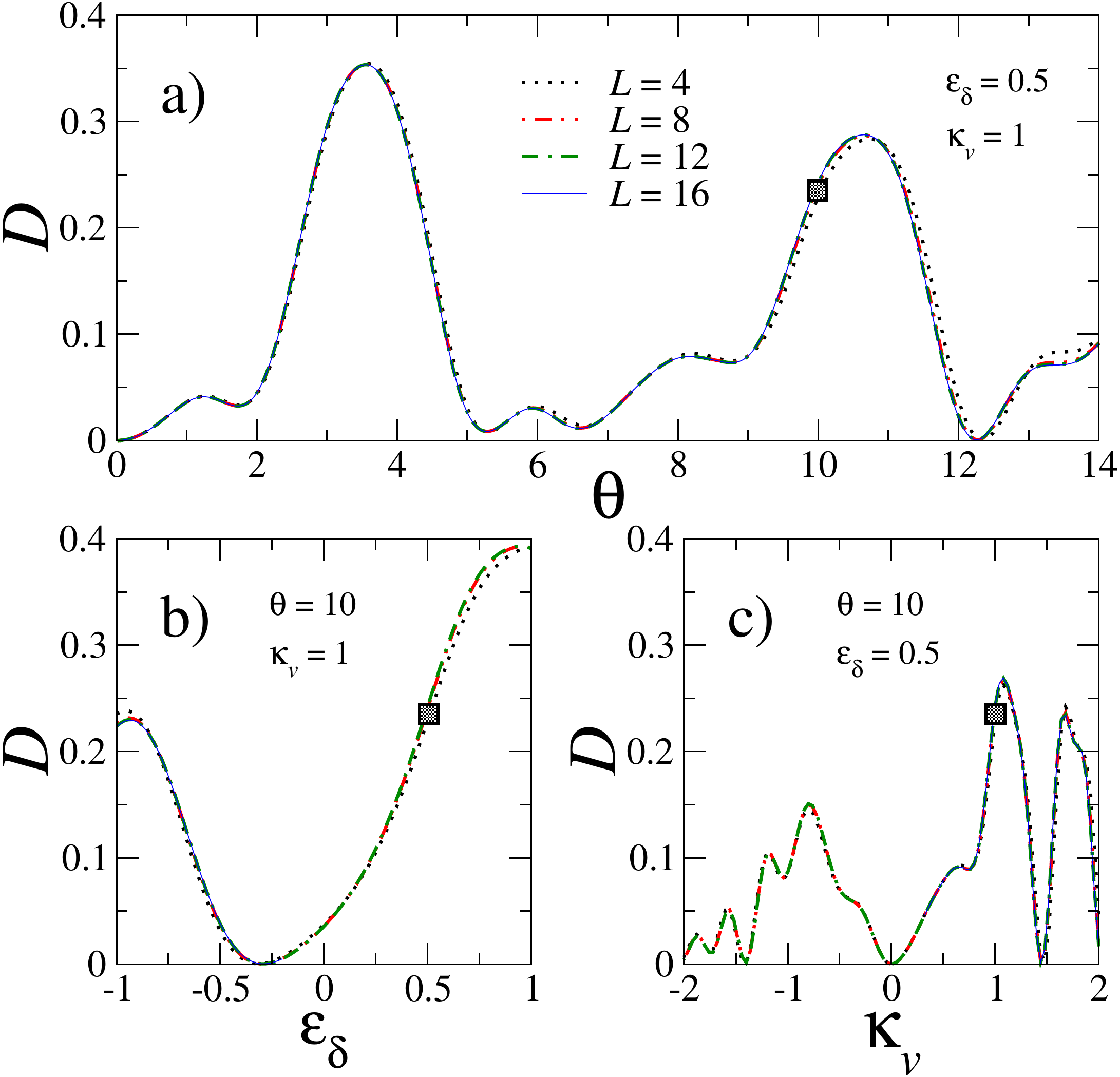}
  \caption{Decoherence function $D$ for a qubit coupled to Ising
    spin-chain systems of different lengths $L$ (see legend) at the
    CQT ($g=g_c=1$).  All numerical data presented here, and in the
    following figures, are for a qubit-system coupling realized
    through $u=0$ and $v \neq 0$ in Eq.~\eqref{hqdef}, such that $[H_q
      ,H_{qS}] \neq 0$.  The three panels display the behavior of $D$
    as a function of various scaling variables, according to the
    following scheme: (a) $\varepsilon_\delta=0.5$, $\kappa_v=1$,
    varying $\theta$; (b) $\kappa_v=1$, $\theta = 10$, varying
    $\varepsilon_\delta$; (c) $\varepsilon_\delta = 0.5$, $\theta =
    10$, varying $\kappa_v$.  In all simulations we fixed
    $\kappa_h=0.8$ and $u=0$, while the qubit was initialized with
    $c_+=\sqrt{2/3}$.  To facilitate the readability, we kept the same
    scaling variables in all the panels, specified in the legend,
    except the one entering the $x$ axis.  The dark square in each
    panel corresponds to a common point in all plots.}
  \label{Purity_CQT}
\end{figure}
%%%%%%%%%%%%%%%%%%%%%%%%%%%%%%%%%%%%%%%%%%%%%%%%%%%%%%%%%%%%%%%%%%%

We start with the analysis of the qubit decoherence function $D$
defined in Eq.~\eqref{rhoq}. Figure~\ref{Purity_CQT} displays the
scaling behavior of $D$ as a function of several different scaling
variables, namely the rescaled time $\theta$ [panel a)], the qubit
detuning $\varepsilon_\delta$ [panel b)], and the qubit-system
coupling $\kappa_v$ [panel c)].  Remarkably, data collapse appears
already for Ising-chain systems of $L \lesssim 10$ sites, as is
evident from the figure.  This validates the ansatz put forward in
Eq.~(\ref{calpscalnc}).  We note that the strongly oscillating
behavior which emerges as a function of $\theta$, implying nearly
perfect revivals of the coherence at short times (e.g., at $\theta
\approx 5.3, \, 6.6, \, 12.3$ in the figure), is due to the fact that
the dynamic FSS framework is probing the post-quench dynamics of a
qubit coupled to a many-body system within the critical regime of a
quantum transition.  In particular, for a fixed value of the rescaled
qubit-system coupling (e.g., $\kappa_v = 1$ in the figure) the
corresponding coupling parameter $v$ entering $H_{qS}$ scales to zero
polynomially with $L$ [recall Eqs.~\eqref{kappauv}
  and~\eqref{kappah}]. Therefore, as observed in Refs.~\cite{PRV-18b,
  V-18}, one expects the global system not to thermalize, and thus a
complete decoherence of the qubit not to occur, for finite and fixed
rescaled coupling constants.

%%%%%%%%%%%%%%%%%%%%%%%%%%%%%%%%%%%%%%%%%%%%%%%%%%%%%%%%%%%%%%%%%%%
\begin{figure}[!t]
  \includegraphics[width=0.95\columnwidth]{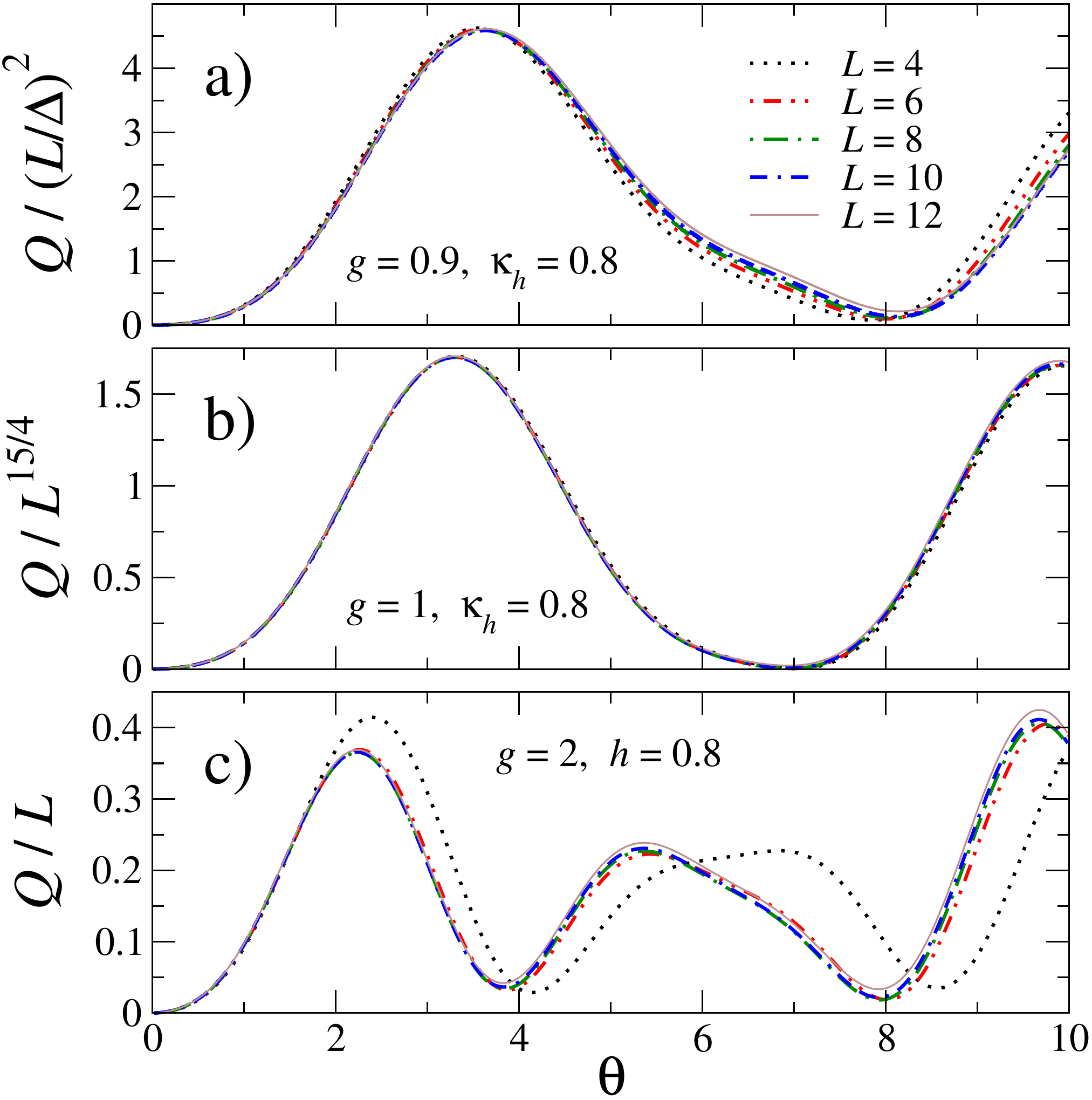}
  \caption{Scaling of the growth-rate function $Q$ as a function of
    time, for three distinct situations: on the FOQT line, for $g=0.9$
    (a); at the CQT, for $g=g_c=1$ (b); in the disordered phase, for
    $g=2$ (c).  The evaluated growth-rate function quantifies the
    sensitivity of the qubit coherence to the coupling $v$. We also
    fixed $\varepsilon_\delta = 0.5$ and set the initial state as $c_+
    = \sqrt{2/3}$.  The longitudinal field $h$ has been chosen without
    loss of generality, in such a way that $\kappa_h = 0.8$ for panels
    (a) and (b), while $h=0.8$ in panel (c).}
  \label{GrowthRate}
\end{figure}
%%%%%%%%%%%%%%%%%%%%%%%%%%%%%%%%%%%%%%%%%%%%%%%%%%%%%%%%%%%%%%%%%%%

A similar reasoning can be drawn for the decoherence growth-rate
function $Q$, which is defined as the second derivative of $D$ with
respect to the qubit-system coupling parameter $u$ or $v$,
cf. Eqs.~(\ref{calqdef}) and (\ref{calqdefv})
respectively. Specifically, we have numerically computed the second
derivative of $D$ with respect to $v$, through the evaluation of
finite differences obtained by varying $\kappa_v$ around zero of a
small step $\delta \kappa_v = \pm 10^{-3}$ (results are stable to the
choice of $\delta \kappa_v$ around such value).  The numerical
outcomes of Fig.~\ref{GrowthRate} display the temporal behavior of $Q$
in three different situations. Namely at a FOQT [panel a),
  cf. Eq.~\eqref{claQscalfoqt}], at the CQT point [panel b),
  cf. Eq.~\eqref{claQscalcqt}], and in the disordered phase [panel c),
  cf. Eq.~\eqref{claQscalnocri}].  A direct comparison of the various
scaling behaviors, which again present a notable data collapse at
small sizes, reveals a dependence on the chain length $L$ which turns
from exponential (at the FOQT, $g<1$) to power law $\propto L^{15/4}$
(at the CQT, $g=g_c=1$) or $\propto L$ (in the disordered phase,
$g>1$).

%%%%%%%%%%%%%%%%%%%%%%%%%%%%%%%%%%%%%%%%%%%%%%%%%%%%%%%%%%%%%%%%%%%
\begin{figure}[!t]
  \includegraphics[width=0.95\columnwidth]{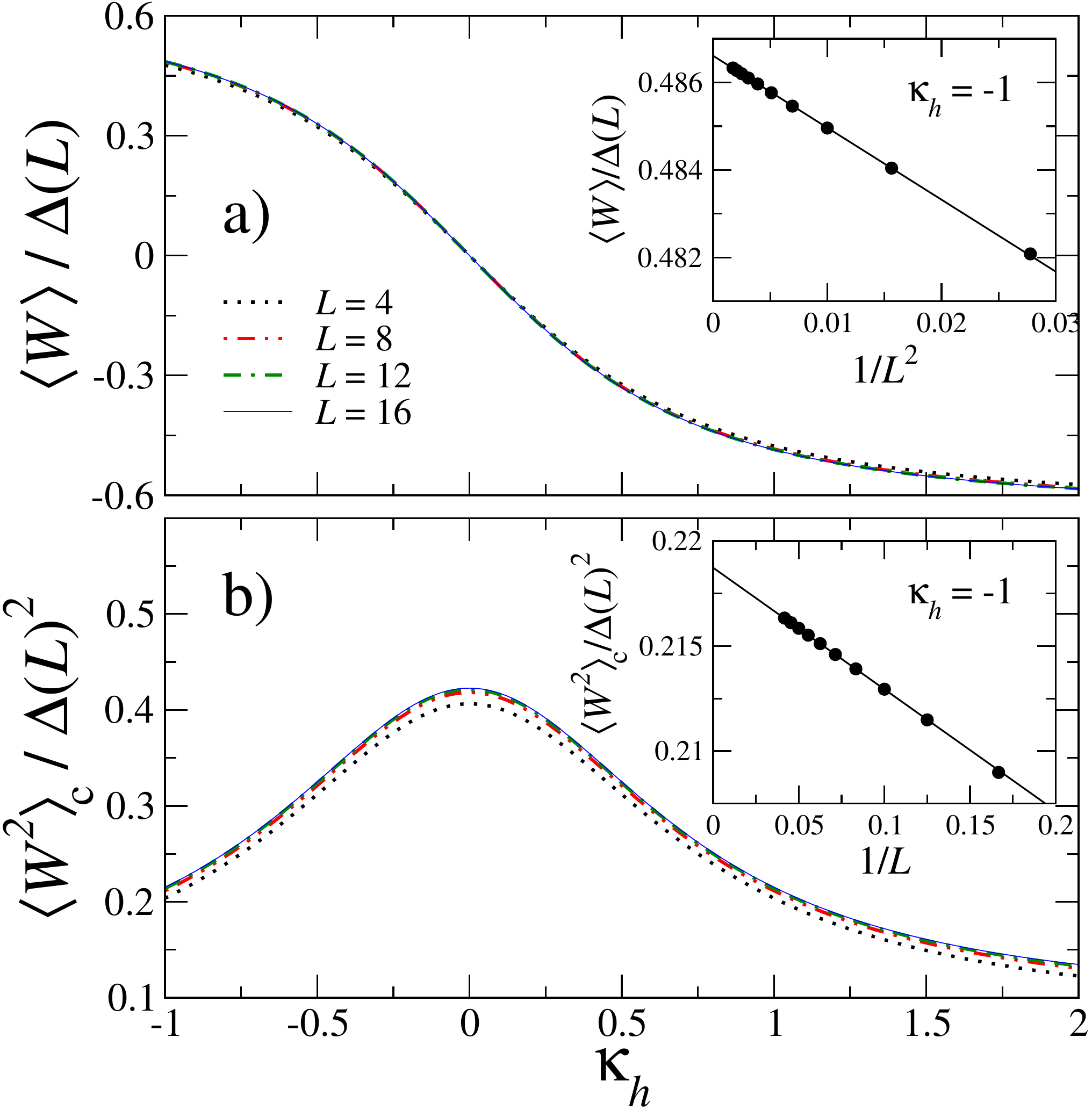}
  \caption{Scaling behavior of the average work $\langle W \rangle$
    [panel a)] and of the variance of the work distribution $\langle
    W^2 \rangle_c = \langle W^2 \rangle - \langle W \rangle^2$ [panel
      b)], done by quenching the qubit-system interaction from zero to
    $\kappa_v=1$, as a function of the rescaled longitudinal field
    $\kappa_h$. The system (Ising chain) is at the CQT point
    $g=g_c=1$, $\varepsilon_\delta = 0.5$, while $c_+ = \sqrt{2/3}$.
    The two insets show the convergence with the system size up to
    $L=24$, at fixed $\kappa_h = -1$, of the average and of the
    variance of the work, respectively showing dominant contributions
    in $L^{-2}$ and $L^{-1}$ (continuous lines are fits of numerical
    data).}
  \label{Work_CQT}
\end{figure}
%%%%%%%%%%%%%%%%%%%%%%%%%%%%%%%%%%%%%%%%%%%%%%%%%%%%%%%%%%%%%%%%%%%

Other properties that we analyzed are related to the statistics
associated with the energy injected by quenching the interaction
strength $H_{qS}$, and with the energy distribution of the system $S$
during the time evolution.  In Fig.~\ref{Work_CQT} we show the average
work $\langle W \rangle$ [panel a)] and the variance of the work
distribution $\langle W^2\rangle_c = \langle W^2 \rangle - \langle W
\rangle^2$ [panel b)] done by the quench in the qubit-system setup, as
a function of the rescaled longitudinal field $\kappa_h$ in the
Ising-chain system at its CQT, for fixed $\epsilon_\delta$.  In
analogy with the decoherence properties of the qubit, numerical
results indicate a nice scaling behavior, thus confirming the dynamic
FSS ansatzes of Eqs.~\eqref{scaloverw_bis}. A closer look at
finite-size corrections for fixed $\kappa_h$ reveals an approach to
the asymptotic behavior which is characterized by $O(L^{-2})$ and
$O(L^{-1}$) corrections, respectively (see the two insets).  This
reflects a slower approach to the expected asymptotic behavior of
$\langle W^2 \rangle_c$, rather than that of $\langle W \rangle$, as
is qualitatively visible by comparing the two main panels.  These data
suggest that the global convergence of full work statistics to its
dynamic FSS may be $O(L^{-1})$, as already pointed out in
Ref.~\cite{NRV-18}.

\begin{figure}[!t]
  \includegraphics[width=0.95\columnwidth]{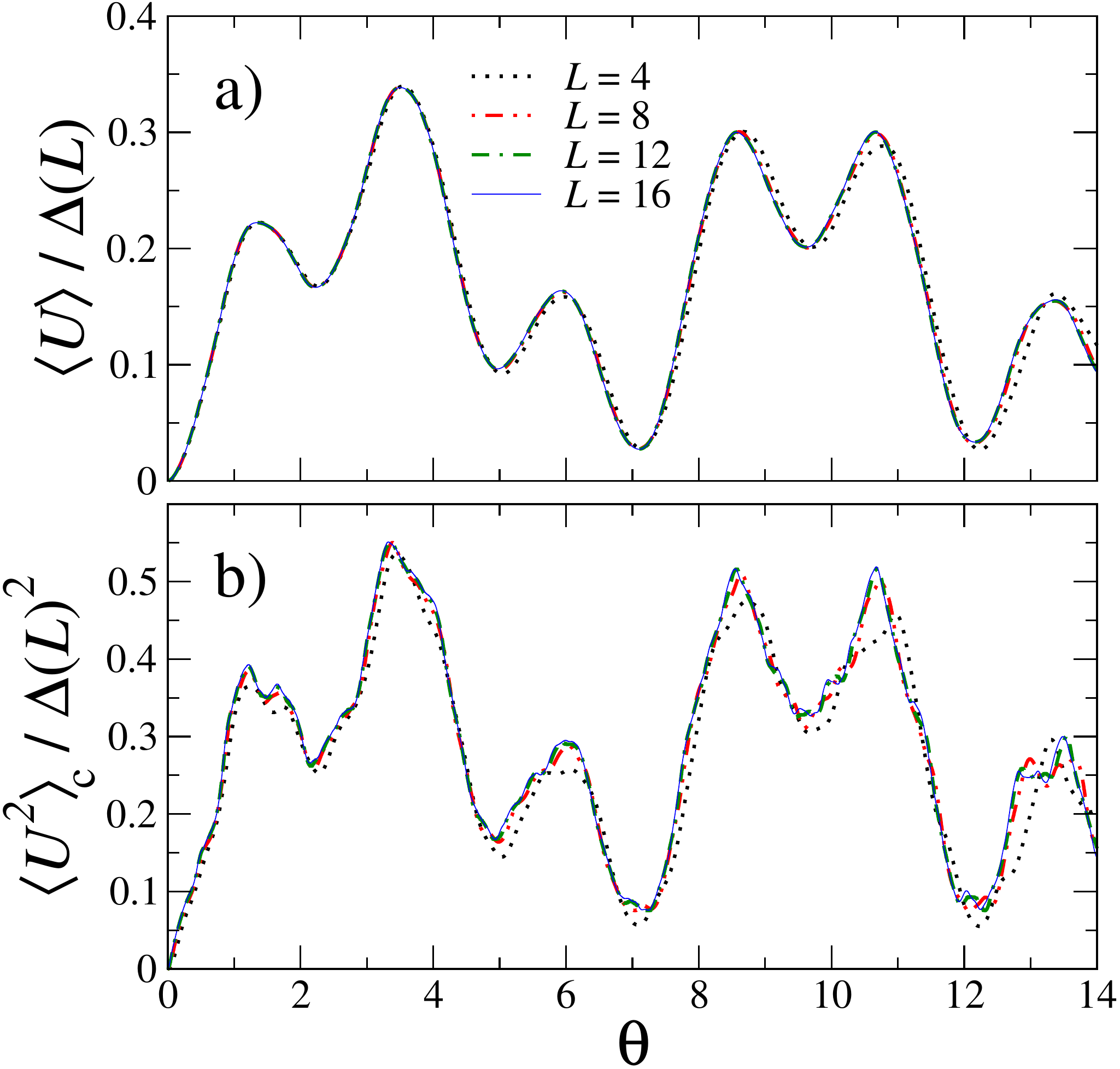}
  \caption{Scaling behavior in time of the energy-difference
    distribution of an Ising-chain system $S$ at the CQT, coupled to a
    qubit.  Panel a) refers to the average energy difference $\langle
    U \rangle$, while panel b) to its fluctuations $\langle U^2
    \rangle_c = \langle U^2 \rangle - \langle U \rangle^2$.  The other
    scaling variables and the initial state have been set as in panel
    a) of Fig.~\ref{Purity_CQT}.}
  \label{EnergyS_CQT}
\end{figure}

In an analogous spirit, it is eventually interesting to analyze the
first two moments of the energy-difference distribution of the Ising
system $S$ along the dynamics at the CQT, cf. Eq.~\eqref{pdedef}.
Specifically, the behavior with respect to the rescaled time $\theta$
of the average energy difference $\langle U \rangle$ and of its
fluctuations $\langle U^2 \rangle_c = \langle U^2 \rangle - \langle U
\rangle^2$ are shown in panels a) and b) of Fig.~\ref{EnergyS_CQT},
respectively.  We observe likewise a nice convergence to the dynamic
FSS behavior predicted by Eqs.~\eqref{escal}. We shall observe that
the calculation of fluctuations is very sensitive to the numerical
accuracy of the simulated dynamics, essentially because much larger
precision is required when computing connected quantities, generally
arising from large cancellations of their terms.

We point out that the numerical results presented in this section
have been obtained by fixing the same initial state and adopting
specific values for the various scaling variables (for details, see the insets
of the various figures). However we have also performed simulations using
other sets of parameters (not shown) and carefully checked that
all our conclusions are not affected by such choices.

%%%%%%%%%%%%%%%%%%%%%%%%%%%%%%%%%%%%%%%%%%%%%%%%%%%%%%%%%%%%%%%%%%%
%%%%%%%%%%%%%%%%%%%%%%%%%%%%%%%%%%%%%%%%%%%%%%%%%%%%%%%%%%%%%%%%%%%

\section{Dynamic behavior at the FOQT}
\label{numresFOQT}

\subsection{Two-level reduction of the system $S$}
\label{sec:2lev-red}

We now turn to a situation where the Ising-chain system $S$ is along
the FOQT line ($g<g_c$), and concentrate on boundary conditions which
do not favor any particular phase.  In such case, one can try to
simplify the description of the global system by employing a two-level
approximation for the system $S$, as was done in Refs.~\cite{PRV-18,
  PRV-18b}, in different contexts.  Under the assumption that only the
lowest levels of the system $S$ are effectively involved by the
dynamic behavior arising from the sudden quench of the qubit-system
interaction in the dynamic FSS limit, we may consider the following
two-level reduction of the Hamiltonian $H_S$:
\begin{eqnarray}
  &&H_{S_2}(h) = - {\beta\over 2}\,\sigma^{(3)} 
  + {\gamma\over 2} \,\sigma^{(1)},\label{Hsred}\\
  &&\beta = 2 m_0 h L^d ,\quad \gamma = \Delta(L),\quad
  \kappa_h = \beta/\gamma\,,
  \nonumber
\end{eqnarray}
which acts on two-component wave functions, corresponding to the
states $|+ \rangle$ and $|-\rangle$, such that
$\sigma^{(3)}|\pm\rangle = \pm |\pm\rangle $.  Then, the qubit-system
interaction term becomes
\begin{eqnarray}
  &&H_{qS_2} = - {\zeta\over 2} \,\Sigma^{(1)} \,\sigma^{(3)} 
  - {\eta\over 2} \,\Sigma^{(3)}\,\sigma^{(3)}\,,
  \label{hqsred}\\
  &&\eta = 2 m_0 L^d u,\qquad \kappa_u = \eta/\gamma\,,\nonumber\\
  &&\zeta = 2 m_0 L^d v,\qquad \kappa_v = \zeta/\gamma\,.\nonumber
\end{eqnarray}
The above Hamiltonian terms are completed by the qubit Hamiltonian,
which we can be written as
\begin{equation}
  H_q = \tfrac12 \delta \Sigma^{(3)},\qquad \varepsilon_\delta
  =\delta/\gamma\,.
  \label{hqrw}
\end{equation}
neglecting the irrelevant identity term of Eq.~(\ref{hqdef}).
Therefore, in this approximation, the global Hamiltonian is given by
\begin{equation}
H_2 = H_{S_2} + H_q + H_{qS_2}\,,
\label{h2level}
\end{equation}
to be compared with Eq.~\eqref{hlamu}.

Within the two-level approximation for the system $S$ at a FOQT, we
may thus write the global Hamiltonian as
\begin{eqnarray}
  \hat{H}_2 \equiv {H_2\over\gamma}
  &=&  - {\kappa_h\over 2} \,\sigma^{(3)} + {1\over 2} \sigma^{(1)} 
  + {\varepsilon_\delta\over 2} \,\Sigma^{(3)}  \label{Hgl2}\\
  &&- {\kappa_u\over 2} \,\Sigma^{(3)} \sigma^{(3)}  
  - {\kappa_v\over 2} \,\Sigma^{(1)} \sigma^{(3)} \,,
  \nonumber
\end{eqnarray}
or, using the bases where both $\Sigma^{(3)}$ and $\sigma^{(3)}$ are
diagonal over the qubit and $S$ states, we may write $\hat{H}_2$ as
the following $4 \times 4$ matrix:
\begin{equation}
{1\over 2} \left( 
\begin{array}{c@{\ \ }c@{\ \ }c@{\ \ }c@{\ \ }}
\varepsilon_\delta - \kappa_{h+u} & 1 
& - \kappa_v & 0\\
1  & \varepsilon_\delta + \kappa_{h+u} 
& 0 & \kappa_v\\
- \kappa_v & 0 
& -\varepsilon_\delta -\kappa_{h-u} & 1\\
0 &\kappa_v 
& 1 & -\varepsilon_\delta + \kappa_{h-u}\\
\end{array}
\right)
\label{h2snonco}
\end{equation}
with $\kappa_{h\pm u} \equiv \kappa_h \pm \kappa_u$.  Note that the
Hamiltonian (\ref{Hgl2}) allows us to write the corresponding
Schr\"odinger problem in terms of scaling variables only, i.e.
\begin{equation}
  i {\partial \over \partial \theta} |\psi(\theta)\rangle =
  \hat{H}_2 |\psi(\theta)\rangle\,, \qquad \theta = \gamma\,t\,,
  \label{sceq2sgl}
\end{equation}
where $|\psi(\theta)\rangle$ denotes the wavefunction of the global
system in the reduced four-dimensional Hilbert space.
Equation~\eqref{sceq2sgl} readily implies that, under the two-level
reduction approximation for the system $S$, the dynamic FSS behavior
put forward in Sec.~\ref{dfss} is automatically guaranteed.

\subsection{Numerical results}
\label{sec:FOQT_num}

We now report a numerical verification of the dynamic FSS behavior
outlined in Sec.~\ref{dfss} within the one-dimensional Ising model
along its FOQT line, and the comparison with the results of the
two-level approximation of the system $S$.

We start commenting on the simpler case $u \neq 0$ and $v=0$, where
the qubit Hamiltonian commutes with the interaction term.  In such
case, one can compute the corresponding FSS functions in an analytic
form.  Indeed the matrix representation~\eqref{h2snonco} of $\hat H_2$
reduces to a $2 \times 2$ block diagonal form, for which analytic
expressions can be obtained. Appendix~\ref{v0case} reports the dynamic
FSS functions of all the quantities defined in Eq.~\ref{definitions}.

The analytic calculations for the less trivial case, $u=0$ and $v\neq
0$, for which $[H_q,H_{qS_2}]\neq 0$, are more cumbersome.  Indeed, the
solution of the corresponding quantum problem requires the
diagonalization of the full $4 \times 4$ matrix
Hamiltonian~\eqref{h2snonco} over the four-dimensional Hilbert space
of the qubit and two levels associated with $S$.  The results are not
very illuminating, and for this reason we decided not to report them
here. Conversely, we preferred to concentrate on a quantitative
comparison between the outcomes of numerical exact diagonalization
simulations for the full many-body Hamiltonian $H$ and those of the
reduced $4 \times 4$ setup $\hat H_2$.

%%%%%%%%%%%%%%%%%%%%%%%%%%%%%%%%%%%%%%%%%%%%%%%%%%%%%%%%%%%%%%%%%%%
\begin{figure}[!t]
  \includegraphics[width=0.95\columnwidth]{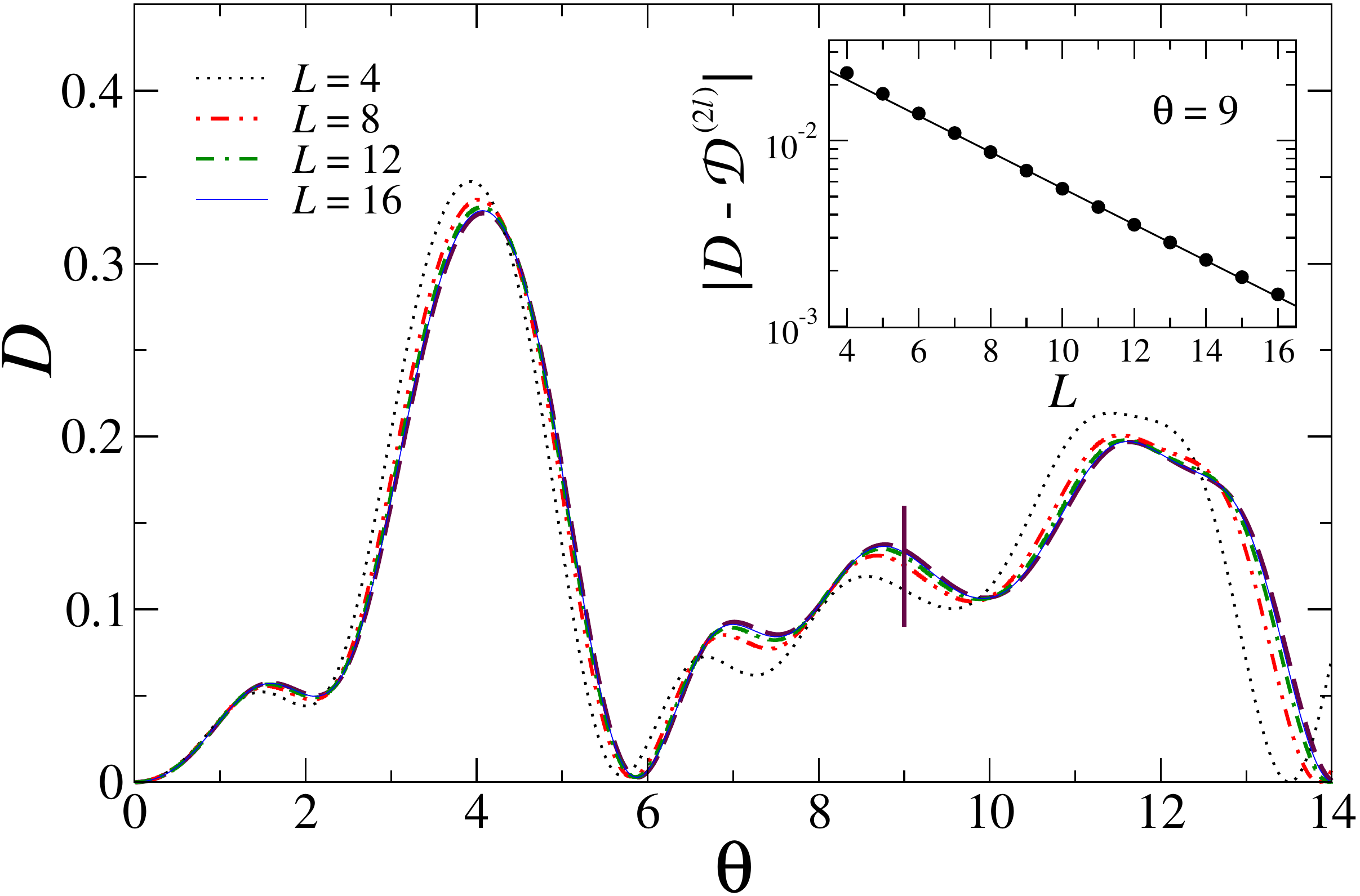}
  \caption{Decoherence function $D$ for a qubit coupled to an Ising
    spin chain at the FOQT ($g=0.9$), as a function of the rescaled
    time $\theta$.  We fixed $\kappa_v=1$, $\varepsilon_\delta = 0.5$,
    and set the initial state with $c_+=\sqrt{2/3}$, as in panel a) of
    Fig.~\ref{Purity_CQT}.  The inset shows numerical data for
    $\theta=9$, supporting an exponential convergence to the
    prediction ${\cal D}^{(2l)}$ given by the two-level approximation
    for the Ising chain (thick dashed line in the main panel).}
  \label{Purity_FOQT}
\end{figure}
%%%%%%%%%%%%%%%%%%%%%%%%%%%%%%%%%%%%%%%%%%%%%%%%%%%%%%%%%%%%%%%%%%%

The analysis of the dynamic FSS for the decoherence function $D$ as a
function of the rescaled time $\theta$, when coupled to the Ising spin
chain at the FOQT, is reported in Fig.~\ref{Purity_FOQT}, where we
plotted the outcomes of the simulation of the full model.  They
demonstrate a clear qualitative accordance with those of the two-level
approximation ${\cal D}^{(2l)}(\theta)$, obtained by solving
Eq.~\eqref{sceq2sgl} (thick dashed brown line).  We notice the
appearance of strong revivals, as is the case in proximity of a CQT,
cf. Fig.~\ref{Purity_CQT}. The convergence in $L$ to the analytic
two-level approximation appears to be exponential, as shown by the
inset of Fig.~\ref{Purity_FOQT} for a fixed value of $\theta =9$.

%%%%%%%%%%%%%%%%%%%%%%%%%%%%%%%%%%%%%%%%%%%%%%%%%%%%%%%%%%%%%%%%%%%
\begin{figure}[!t]
  \includegraphics[width=0.95\columnwidth]{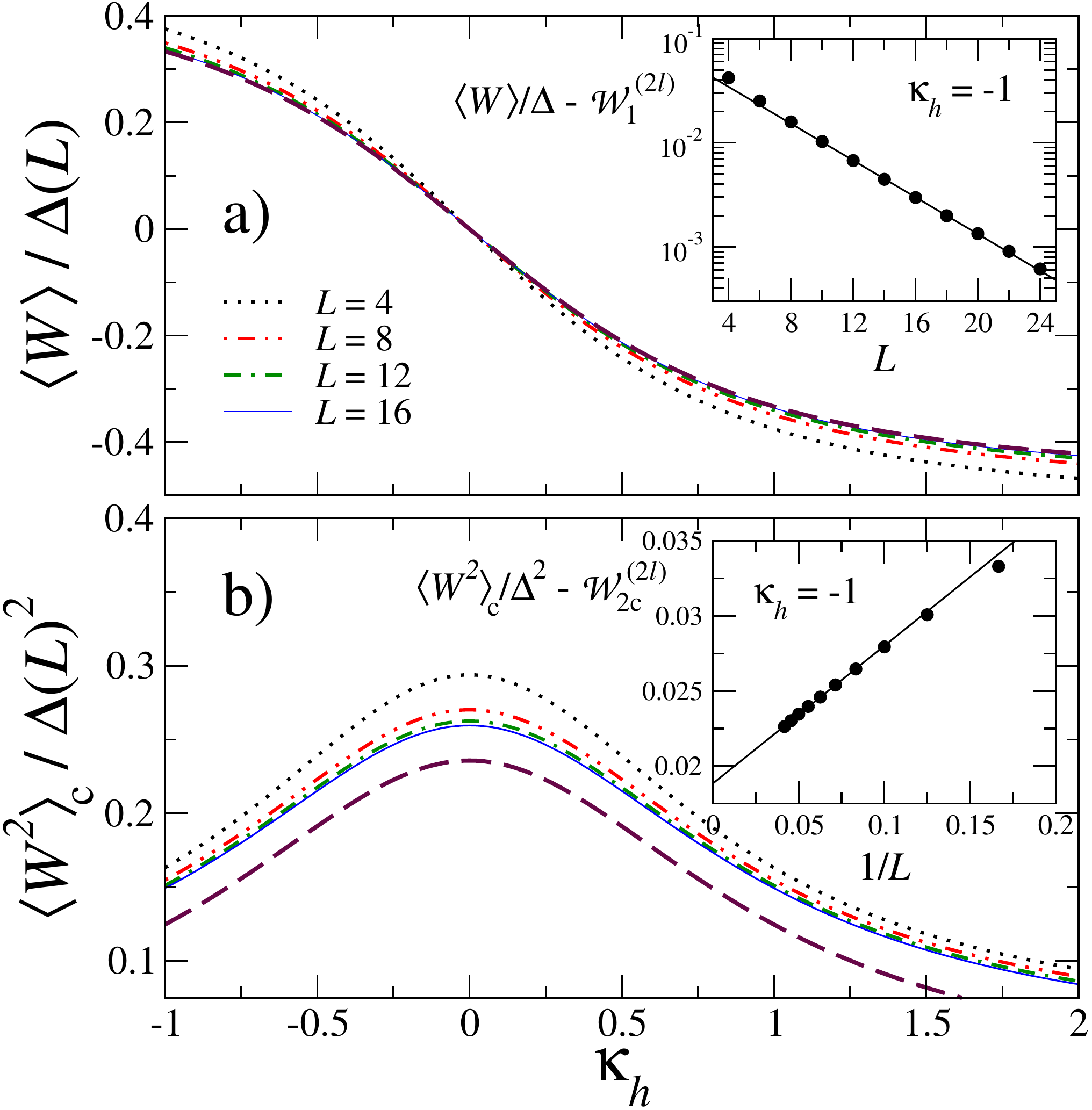}
  \caption{Same as in Fig.~\ref{Work_CQT}, but for a qubit coupled to
    an Ising spin-chain system $S$ at the FOQT, with $g=0.9$.  Panels
    a) and b) show curves for the average work and its fluctuations,
    respectively.  Thick dashed lines denote the predictions, ${\cal
      W}_1^{(2l)}$ and ${\cal W}_{2c}^{(2l)}$ respectively, given by
    replacing $S$ with an approximate two-level model. The inset of
    panel a) displays an exponential convergence in $L$ of the ratio
    $\langle W \rangle / \Delta(L)$ to the two-level prediction ${\cal
      W}_1^{(2l)}$. The inset of panel b) shows the behavior in $L$ of
    the difference between the work fluctuations $\langle W^2
    \rangle_c / \Delta(L)^2$ and the two-level prediction ${\cal
      W}_{2c}^{(2l)}$. In this case, a $L^{-1}$ fit of the numerical
    data leads to a finite discrepancy of $1.89(1) \times 10^{-2}$ in
    the infinite-volume limit.  In both insets we fixed $\kappa_h =
    -1$.}
  \label{Work_FOQT}
\end{figure}
%%%%%%%%%%%%%%%%%%%%%%%%%%%%%%%%%%%%%%%%%%%%%%%%%%%%%%%%%%%%%%%%%%%

We now look at the statistics of energy exchanges.  The first two
moments of the statistics of the work, for the Ising-chain system at
the FOQT, are reported in Fig.~\ref{Work_FOQT}.  Data collapse to a
dynamic FSS behavior in the infinite volume volume is clearly
evident. However, while the average work $\langle W \rangle$ nicely
converges to the two-level prediction ${\cal W}^{(2l)}$, with an
apparent exponential dependence on $L$ [panel a) and its inset], this
is not the case for the work fluctuations $\langle W^2 \rangle_c$
[panel b)].  Specifically, the two-level prediction ${\cal
  W}_{2c}^{(2l)}$ is apparently off from the expected limiting
behavior.  In fact, the inset hints at a $O(L^{-1})$ convergence of
the discrepancy to a value which is different from zero, thus implying
the failure of the two-level reduction of the system $S$ in exactly
grasping the asymptotic behavior of the higher momenta of the work
statistics.  Further details on the accuracy of the two-level
approximation are given in App.~\ref{acc2lev}.

%%%%%%%%%%%%%%%%%%%%%%%%%%%%%%%%%%%%%%%%%%%%%%%%%%%%%%%%%%%%%%%%%%%
\begin{figure}[!t]
  \includegraphics[width=0.95\columnwidth]{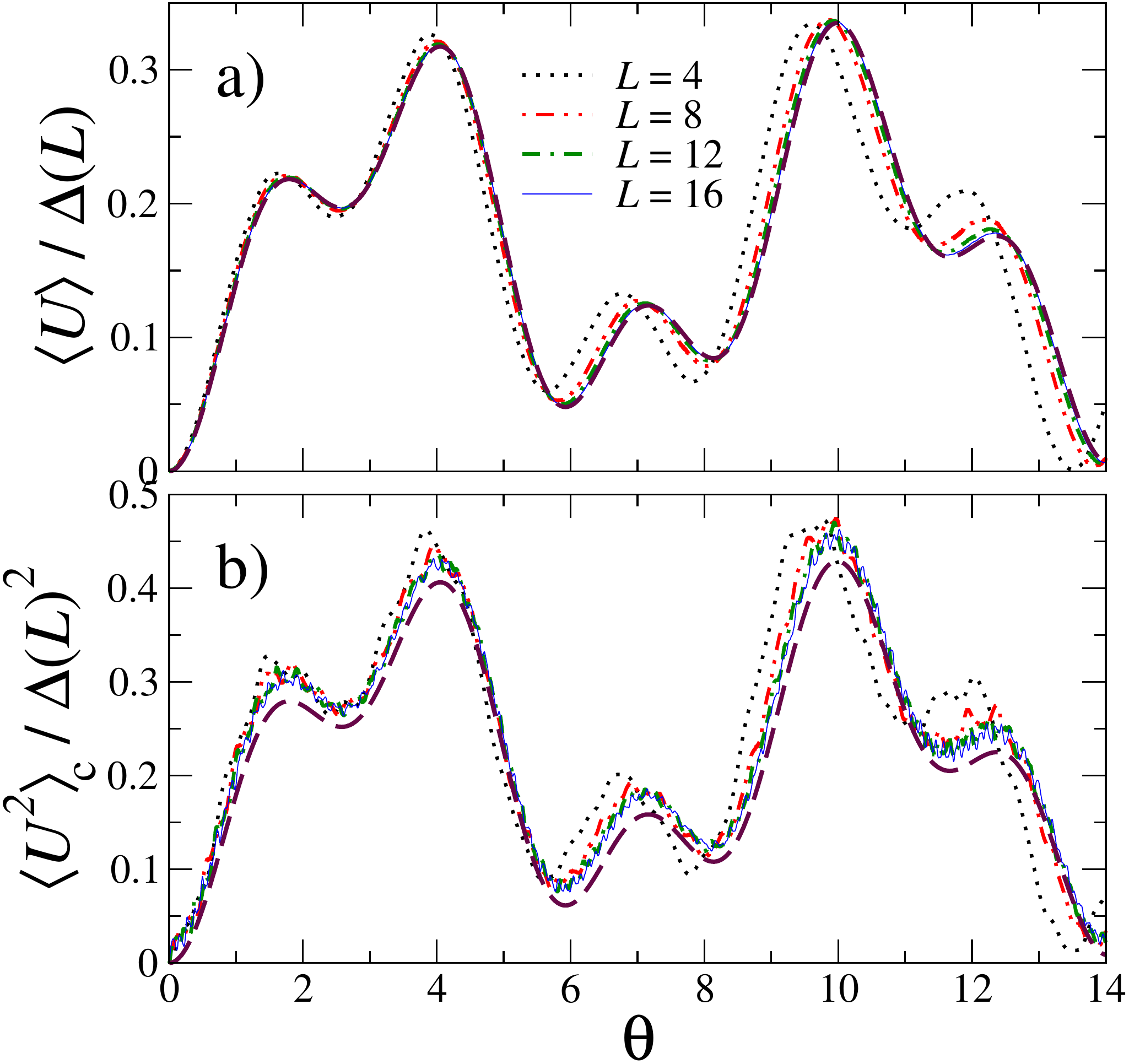}
  \caption{Same as in Fig.~\ref{EnergyS_CQT}, but for a qubit coupled
    to an Ising-chain system at the FOQT, with $g=0.9$.  Thick dashed
    lines denote the predictions, ${\cal U}^{(2l)}_1$ and ${\cal
      U}^{(2l)}_2$ respectively, given by replacing $S$ with an
    approximate two-level model as depicted in
    Sec.~\ref{sec:2lev-red}.  A small Runge-Kutta time step of
    $dt=10^{-4}$ was employed, in order to guarantee the numerical
    accuracy of the results plotted in panel b).}
  \label{EnergyS_FOQT}
\end{figure}
%%%%%%%%%%%%%%%%%%%%%%%%%%%%%%%%%%%%%%%%%%%%%%%%%%%%%%%%%%%%%%%%%%%

Finally, in Fig.~\ref{EnergyS_FOQT} we repeat the analysis of the
statistics of the energy-difference distribution in the system $S$,
where similar conclusions apply.  In particular, we display the
temporal behavior of the average [panel a)] and of its fluctuations
[panel b)].  Analogously to the outcomes we found at the CQT
(cf. Fig.~\ref{EnergyS_CQT}), we observe that fluctuations $\langle
U^2 \rangle_c$ are very sensitive to the accuracy of the simulation.  It
is also worth noticing that the thick dashed curves which indicate the
two-level predictions (${\cal U}_1^{(2l)}$ and ${\cal U}_2^{(2l)}$,
respectively) follow the same behavior with $\theta$. More precisely,
it can be shown that the ratio
\begin{equation}
  {\cal U}_2^{(2l)} / {\cal U}_1^{(2l)} = \sqrt{1 + \kappa_h^2} 
\end{equation}
depends only on the value of $\kappa_h$ (in the figure we used
$\kappa_h = 0.8$, therefore $\sqrt{1 + \kappa_h^2} \approx 1.28$).
This comes in analogy with Eq.~\eqref{ersquare}, which can be easily
proven for a longitudinal qubit-system interaction ($v=0$).

Before ending we stress that, similarly to the dynamic behavior at the CQT,
analogous results have been obtained (and, in particular, the collapse
of numerical data to the FSS behavior put forward in Sec.~\ref{scalvarfoqt})
for other values of scaling variables and initial states.

\section{Conclusions}
\label{sec:concl}

We have addressed the quantum dynamics of a system composed of a qubit
globally coupled to a many-body system characterized by short-range
interactions.  We employed a dynamic FSS framework to investigate the
out-of-equilibrium dynamics arising from the sudden variation (turning
on) of the interaction between the qubit and the many-body system, in
particular when the latter is in proximity of a quantum first-order or
a continuous phase transition.  Although the approach is quite
general, we considered $d$-dimensional quantum Ising spin models in
the presence of transverse and longitudinal fields, as paradigmatic
quantum many-body systems.

To characterize the out-of-equilibrium dynamics, we focused on a
number of quantum-information oriented properties of the model.
Generalizing the results of Ref.~\cite{V-18}, we considered the
information and energy flow among the various parts of the composite
system: we studied the decoherence of the qubit and the statistics
associated with the energy injected by switching on the qubit-system
interaction and with the energy distribution of the system during the
temporal evolution.  When the many-body system $S$ is close at a
quantum transition, either at a CQT or at a FOQT, we derived the
asymptotic scaling behavior exploiting the dynamic FSS framework.  The
scaling behaviors of the above quantities were validated by means of
extensive numerical simulations specialized to one-dimensional Ising
systems.  We always observed convergence to the expected asymptotic
FSS behavior when the system $S$ is at both CQTs and FOQTs.

In the case of FOQTs, we also employed a two-level approximation for
the many-body system $S$, to compute the dynamic FSS functions
associated with the out-of-equilibrium quantum evolution.  The
agreement with the numerical results is satisfactory.  Quantitative
differences between the numerics and the two-level approximation
emerge only when monitoring fluctuations and higher momenta of the
various energy statistics.

It is however worth pointing out that, already with ten spins, it is
possible to infer the asymptotic FSS behavior with a fair accuracy.
This paves the way toward an experimental probe of the influence of
criticality on the quantum transport properties, spurred on by the
recent developments in quantum technologies with ultracold atoms and
ions.  Indeed they already demonstrated the capability to faithfully
reproduce the unitary dynamics of quantum Ising-like chains with
$\approx 10$ spins~\cite{Simon-etal-11, Edwards-etal-10,
  Islam-etal-11, LMD-11, Kim-etal-11, Debnath-etal-16,
  Labuhn-etal-16}.

Finally we mention that it is possible to extend our dynamic FSS
analysis to more general situations.  Beside those based on sudden
variations of the interactions between the qubit and the many-body
system, one may consider other dynamic protocols, for example by
taking the opposite limit of slow changes. This situation can be also
analyzed within appropriate FSS frameworks, such as that considered in
Ref.~\cite{PRV-18}.  We may also devise extensions to more general
models, where the qubit is replaced by a generic $N$-level quantum
system, the environment is mapped in the continuum limit (or, more
generally, can be modeled by a many-body system presenting CQTs or
FOQTs), and the qubit-system coupling is not homogeneous or may have a
different and more complicated shape.

\appendix

\section{The commutative case $[H_q,H_{qS}]=0$}
\label{commcase}

A particular case of the general problem outlined in Sec.~\ref{gset}
is realized when $v=0$ in the qubit-system Hamiltonian~(\ref{hqdef}),
i.e.,
\begin{equation}
  H_{qS} = u \Sigma^{(3)} P \, ,
\end{equation}
which implies $[H_q, H_{qS}]=0$. This condition allows us to write the
time evolution of the global system in terms of dynamic evolutions of
the system $S$ only.  Indeed, one can easily prove that the solution
of the corresponding Schr\"odinger problem is given by
\begin{eqnarray}
  |\Psi(t)\rangle  &=& e^{- i \epsilon_+ t} 
  c_+ | + \rangle \otimes | \Phi_{h+u}(t) \rangle  \nonumber \\
  &+& e^{- i \epsilon_- t} c_- |- \rangle \otimes | \Phi_{h-u}(t) \rangle\,,
  \label{psitcom}
\end{eqnarray}
where 
\begin{equation}
  | \Phi_{h\pm u}(t)\rangle = e^{-iH_S(h\pm u) t} |0_h\rangle\,,
\label{phipm}
\end{equation}
i.e., they are solutions of the Schr\"odinger equations for the system
$S$ only,
\begin{equation}
  i {\partial \over \partial t} |\Phi_{h\pm u}(t)\rangle 
  = H_S(h\pm u) |\Phi_{h \pm u}(t)\rangle\,,
  \label{sceqb}
\end{equation}
with $|\Phi_{h\pm u}(t=0)\rangle = | 0_h\rangle$.

Notable relations can be obtained focusing on the evolution of the
qubit only.  The elements of its reduced density matrix,
cf. Eq.~(\ref{rhoq}), read:
\begin{eqnarray}
&&\rho_{q,11}(t) = |c_+|^2\,,
\quad 
\rho_{q,22}(t) = |c_-|^2\,,
\label{qdmco}\\ 
&&\rho_{q,12}(t) = e^{-i\delta t} c_-^* c_+ 
\langle \Phi_{h-u}(t)|\Phi_{h+u}(t)\rangle = 
\rho_{q,21}(t)^*\,.
\nonumber
\end{eqnarray}
The decoherence function $D(t)$, cf. Eq.~(\ref{ddef}), can be written
as
\begin{eqnarray}
D(t) = 2 |c_+|^2 |c_-|^2 F_D(t)\, ,
\label{purity}
\end{eqnarray}
where
\begin{eqnarray}
F_D(t) = 1 - | \langle \Phi_{h-u}(t) | \Phi_{h+u}(t)
\rangle |^2\,,
\label{Sdef}
\end{eqnarray}
and $0\le F_D(t)\le 1$.  The function $F_D$ measures the quantum
decoherence, quantifying the departure from a pure state.  Indeed
$F_D(t)=0$ implies that the qubit is in a pure state, while $F_D(t)=1$
indicates that the qubit is maximally entangled, corresponding to a
diagonal density matrix
\begin{equation}
\rho_q = {\rm diag} \big[ |c_+|^2,|c_-|^2 \big] \, .
\label{maxentrho}
\end{equation}
Notice that the decoherence functions $D(t)$ and $F_D(t)$ do not
depend on the spectrum of the qubit Hamiltonian, and in particular on
$\delta$.
 
%Note also the relation
%\begin{eqnarray}
%{\rm Re} \, \langle \Psi_{-u}(t) |\Psi_u(t)\rangle = 
%{\rm Re}\, \langle \Phi_{h-u}(t) |\Phi_{h+u}(t)\rangle\,.
%\label{overlap}
%\end{eqnarray}
%between the overlap of time evolutions using coupling $u$ and $-u$.  

We also note that, as a consequence of the commutativity between qubit
Hamiltonian $H_q$ and the interaction term $H_{qS}$, the average qubit
energy $E_q $ does not change along the quantum evolution of the
global system. Indeed its value
\begin{equation}
  E_q = \langle \Psi(t)| H_q |\Psi(t)\rangle = {\rm Tr}\, \big[ \rho_q \,H_q \big]
  = \sum_{i=\pm} \epsilon_\pm |c_\pm|^2
  \label{avHqc}
\end{equation}
remains constant, and therefore it is determined by the initial
condition of the qubit.

Since the average total energy must remain constant during the global
evolution, Eq.~(\ref{avHqc}) also implies that
\begin{equation}
E - E_q = 
\langle \Psi(t)| H_s + H_{qS} |\Psi(t)\rangle 
\label{parene}
\end{equation}
remains constant.  Concerning the average work to perform the quench,
cf. Eq.~(\ref{workdef}), since $E_q$ remains unchanged, simple
calculations lead to
\begin{equation}
  \langle W \rangle = \langle \Psi_0| H_{qS}| \Psi_0\rangle
  = - u\,(|c_+|^2 - |c_-|^2) \, \langle 0_h | P | 0_h \rangle,
  \label{auwork}
\end{equation}
Notice that $\langle W \rangle=0$, when $|c_+|=|c_-|$.

\subsection{Dynamic FSS at FOQTs, for $v=0$}
\label{v0case}

As discussed in Sec.~\ref{numresFOQT}, at FOQTs one may effectively
replace the many-body system $S$ with an approximate two-level model,
such that the composite qubit-system Hamiltonian is written in the
matrix form~\eqref{h2snonco}.  In the specific case $v=0$, the latter
reduces to a block diagonal form, for which analytic expression can be
derived easily.

In practice, the solution of the corresponding Schr\"odinger
problem~\eqref{sceq2sgl} is still formally given by
Eq.~(\ref{psitcom}).  The many-body states $|\Phi_{h\pm
  u}(\theta)\rangle$ are now replaced by the two-level system states
$|\phi_{h\pm u}(\theta)\rangle$, as obtained by solving
\begin{eqnarray}
  i {\partial \over \partial \theta} |\phi_{h\pm u}(\theta)\rangle
  & = & \hat{H}_{S_2}(h\pm u) |\phi_{h\pm u}(\theta)\rangle\,, 
  \label{sceq2s}\\
  \hat{H}_{S_2}(h\pm u) & = & {H_{S_2}(h\pm u)/\gamma}\,,
\end{eqnarray}
with $H_{S_2}(h)$ as in Eq.~\eqref{Hsred}.
The initial condition is given by the ground state
\begin{equation}
  |\phi_{h\pm u}(\theta \! = \! 0)\rangle = |0_h\rangle =
  \sin \! \Big( \! \frac{\alpha_h}{2} \! \Big) |-\rangle
  - \cos \! \Big( \! \frac{\alpha_h}{2} \! \Big) |+\rangle,
  \label{eigstatela0}
\end{equation}
with $\tan \alpha_h = \kappa_h^{-1}$, and $\alpha_h \in (0,\pi)$.  The
quantum evolution described by Eq.~(\ref{sceq2s}) can be easily
obtained by diagonalizing the $2 \times 2$ Hamiltonian $H_{S_2}(h\pm
u)$, whose eigenstates are
\begin{subequations}
  \begin{eqnarray}
    |0_{h\pm u}\rangle & = & \sin \! \Big( \! \frac{\alpha_{h\pm u}}{2} \! \Big) |-\rangle
    - \cos \! \Big( \! \frac{\alpha_{h\pm u}}{2} \! \Big) |+\rangle \, , \qquad \\
    |1_{h\pm u}\rangle & = & \cos \! \Big( \! \frac{\alpha_{h\pm u}}{2} \! \Big) |-\rangle
    + \sin \! \Big( \! \frac{\alpha_{h\pm u}}{2} \! \Big) |+\rangle \, ,
  \end{eqnarray}
  \label{eigstates}%
\end{subequations}
with $\tan \alpha_{h\pm u} = \kappa_{h\pm u}^{-1}$ and $\alpha_{h\pm
  u} \in (0,\pi)$.  The corresponding energy eigenvalues are
\begin{equation}
  E_{0/1} = \Delta(L) \; {\cal E}_{0/1}, \qquad
  {\cal E}_{0/1} = \mp \tfrac12 \sqrt{1 +  \kappa_{h\pm u}^2}\,.
  \label{eeig}
\end{equation}
The time-dependent state evolves as
\begin{eqnarray}
  |\phi_{h\pm u}(\theta)\rangle & = & e^{-i {\cal E}_0 \theta}\,
  \cos \! \Big( {\alpha_h-\alpha_{h\pm u}\over 2} \Big) 
  |0_{h\pm u}\rangle \nonumber \\
  && +  e^{-i {\cal E}_1 \theta}
  \sin \! \Big( {\alpha_h-\alpha_{h\pm u}\over 2} \Big) |1_{h\pm u}\rangle\,. \quad
\label{psitfo}
\end{eqnarray}
Then, by rewriting them in terms of the original basis $|\pm\rangle$,
using Eqs.~\eqref{eigstates}, and replacing into Eq.~(\ref{psitcom}),
with $|\phi_{u \pm u}(\theta)\rangle$ instead of $|\Phi_{u \pm
  u}(\theta)\rangle$, we obtain the solution of the dynamic problem
within the two-level approximation of the system $S$. This reads
\begin{eqnarray}
  |\psi(\theta)\rangle & = & e^{-i(\varepsilon_\delta/2) \theta} c_+ |
  + \rangle \otimes | \phi_{h+u}(\theta) \rangle \nonumber \\ & + &
  e^{+i(\varepsilon_\delta/2) \theta} c_- | - \rangle \otimes |
  \phi_{h-u}(\theta) \rangle \,.
  \label{psitcom2l}
\end{eqnarray}

This solution is already written in terms of the scaling variables,
thus the scaling behaviors put forward in Sec.~\ref{dfss} are fully
confirmed. The corresponding FSS functions can be analytically
computed from their definitions.  The scaling function ${\cal D}$
associated with the decoherence function $D$, cf. Eqs.~(\ref{ddef})
and (\ref{calpscal}), is given by
\begin{equation}
  {\cal D}(\kappa_u,\kappa_h,\theta) = 4 |c_+|^2 |c_-|^2\,\kappa_u^2\,
  \frac{ 1 - {\rm cos}(\theta\sqrt{1+\kappa_h^2})}{(1+\kappa_h^2)^2}\,.
\label{cdanres}
\end{equation}
Coming to the average work defined in Eq.~\eqref{workdef}, whose
expected scaling behavior is reported in Eq.~(\ref{scaloverw}), we
obtain
\begin{equation}
  {\cal W}^{(2l)}_1(\kappa_u,\kappa_h) = 
  - \frac12 (|c_+|^2 - |c_-|^2) \,\kappa_u \bigg[ 1-
    2\sin \! \Big( \frac{\alpha_h}{2} \Big)^{\!2} \, \bigg] . 
  \label{auworktwol}
\end{equation}
On the other hand, for the second moment 
$\langle W^2\rangle$ we simply obtain 
\begin{equation}
  {\cal W}^{(2l)}_2(\kappa_u,\kappa_h) = 
{1\over 4} \kappa_u^2 \,.
  \label{auworktw2ol}
\end{equation}

For the energy fluctuations of the many-body system $S$, we may obtain
the time dependence of the average energy variation,
cf. Eq.~(\ref{deltastpet}), using the formulas
\begin{equation}
{\cal U}_1^{(2l)}(\theta) = {\rm Tr} \big[ \hat{H}_{S_2}(h)
  \rho(\theta) \big] - {\cal E}_{s0}, \qquad {\cal E}_{s0} = -
\tfrac12 \sqrt{1+\kappa_h^2}\,.
\label{toverle}
\end{equation}
Then, using Eq.~(\ref{psitcom2l}), we may write it as
\begin{eqnarray}
  {\cal U}^{(2l)}_1(\theta) &=& |c_+|^2 
  \langle \phi_{h+u}(\theta)|\hat{H}_{S_2}(h)|\phi_{h+u}(\theta)\rangle
  \nonumber \\
  &+& |c_-|^2 \langle \phi_{h-u}(\theta)|\hat{H}_{S_2}(h)|\phi_{h-u}(\theta)\rangle \,,
  \label{overE}
\end{eqnarray}
where 
\begin{equation}
  \hat{H}_{S_2}(h) = -{\kappa_h\over 2}\sigma^{(3)} + {1\over 2} \sigma^{(1)} \,.
  \label{h2shat}
\end{equation}
Finally, for the average of the square energy variation we obtain
\begin{eqnarray}
  {\cal U}^{(2l)}_2(\theta) &=& {\rm Tr} \big\{ [\hat{H}_{S_2}(h)]^2 \rho(\theta) \big\}
  - 2 {\cal E}_{s0} \,{\cal U}_1^{(2l)}(\theta)  - {\cal E}_{s0}^2 \nonumber\\
  &=& \sqrt{1 + \kappa_h^2}\,\, {\cal U}^{(2l)}_1(\theta) \,,
  \label{ersquare}
\end{eqnarray}
where we used the fact that $[\hat{H}_{S_2}(h)]^2 = (1 + \kappa_h^2) I_2$.

\section{Accuracy of the two-level approximation at the FOQT}
\label{acc2lev}

In Sec.~\ref{sec:FOQT_num} we observed that a two-level reduction of
the many-body system to which the qubit is coupled, when the former is
at a FOQT, is capable to accurately grasp the asymptotic FSS behavior
of several properties of the global system, including the decoherence
quantifiers for the qubit, the averages of the work done by the quench
and of the energy pumped in the system $S$.  All these quantities are
linear functionals of the Hamiltonian $H$, for which the adiabatic
theorem typically applies without any issue~\cite{PRV-18}.

%%%%%%%%%%%%%%%%%%%%%%%%%%%%%%%%%%%%%%%%%%%%%%%%%%%%%%%%%%%%%%%%%%%
\begin{figure}[!t]
  \includegraphics[width=0.95\columnwidth]{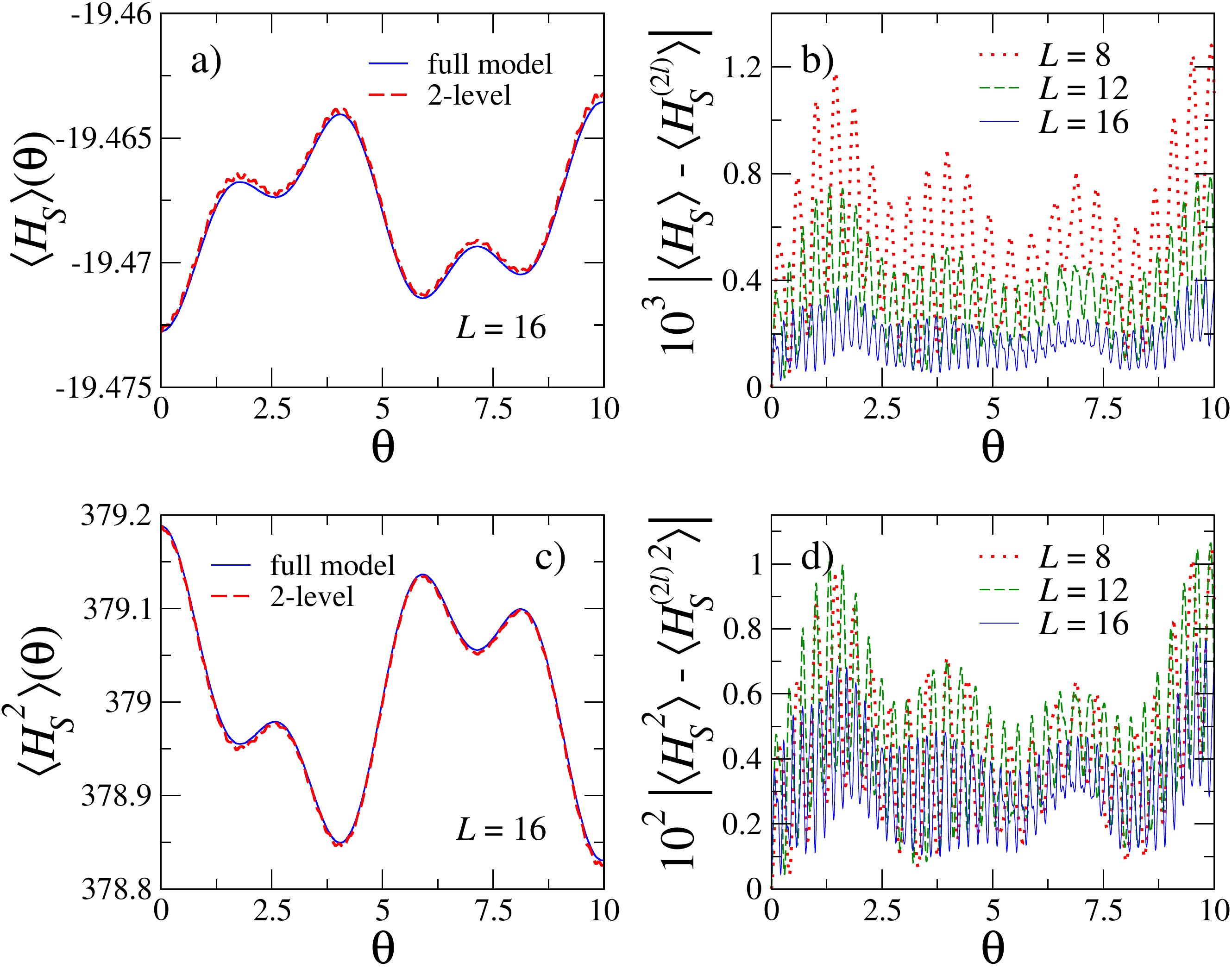}
  \caption{Comparison of the average system energy $\langle H_S
    \rangle$ [panels a) and b)] and of its square value $\langle H_S^2
    \rangle$ [panels c) and d)], evaluated using the full many-body
    Hamiltonian and a truncation to its two lowest levels.  The Ising
    system is at the FOQT with $g=0.9$, while all the other parameters
    are set as in Fig.~\ref{EnergyS_FOQT}.  Panels a) and c) display
    the two evolutions, with respect to the rescaled time $\theta$,
    for a fixed chain length $L = 16$.  Panels b) and d) highlight the
    absolute differences between the two cases, for various system
    sizes.}
  \label{Check_HamS_elem}
\end{figure}
%%%%%%%%%%%%%%%%%%%%%%%%%%%%%%%%%%%%%%%%%%%%%%%%%%%%%%%%%%%%%%%%%%%

On the opposite hand, Figs.~\ref{EnergyS_CQT} and~\ref{EnergyS_FOQT}
have spotlighted clear discrepancies, when comparing fluctuations of
the work and of the system energy (i.e., $\langle W^2\rangle_c$ and
$\langle U^2\rangle_c$), with numerical exact diagonalization results
for the quantum Ising chains.  We believe that such discrepancies are
essentially related to the limited accuracy of the two-level
approximations, which of course cannot capture the full complexity of
a many-body quantum system.  Below we provide evidence of this fact,
focusing on a specific quantity.  It would be tempting to further
investigate the problem, in such a way to achieve a more exhaustive
understanding of the accuracy of our approximation.  This however lies
outside the purposes of this paper and will be left for a future work.

Figure~\ref{Check_HamS_elem} compares the time behavior of the system
energy and of its square value, evaluated either with the full Hamiltonian
$H_S$, or by keeping only its two lowest-energy levels
$|\Phi_1\rangle$ and $|\Phi_2\rangle$ (associated with the two
energies $\varepsilon_1$ and $\varepsilon_2$).  Namely, $\langle
\Psi(\theta)| H_S | \Psi(\theta)\rangle$ or $\langle \Psi(\theta) |
H_S^{(2l)}| \Psi(\theta) \rangle$ respectively, with $H_S^{(2l)} =
\varepsilon_1 |\Phi_1\rangle + \varepsilon_2 |\Phi_2\rangle$.  We
observe a general agreement between the two approaches [panels a) and
  c)], however a more detailed analysis reveals that the discrepancies
among them are typically one order of magnitude larger for $\langle
H_S^2 \rangle$ [panel b)], rather than for $\langle H_S \rangle$
[panel d)].  Moreover, while the data in panel b) suggest that such
discrepancies systematically diminish with increasing the size, the
situation in panel d) is less clear and fluctuations at $L=16$ are
still quite large.
Notice also the appearance of wiggles in panels a) and c), concerning
the results obtained with a two-level truncation of the Hamiltonian
spectrum.

%%%%%%%%%%%%%%%%%%%%%%%%%%%%%%%%%%%%%%%%%%%%%%%%%%%%%%%%%%%%%%%%%%%
\begin{figure}[!t]
  \includegraphics[width=0.95\columnwidth]{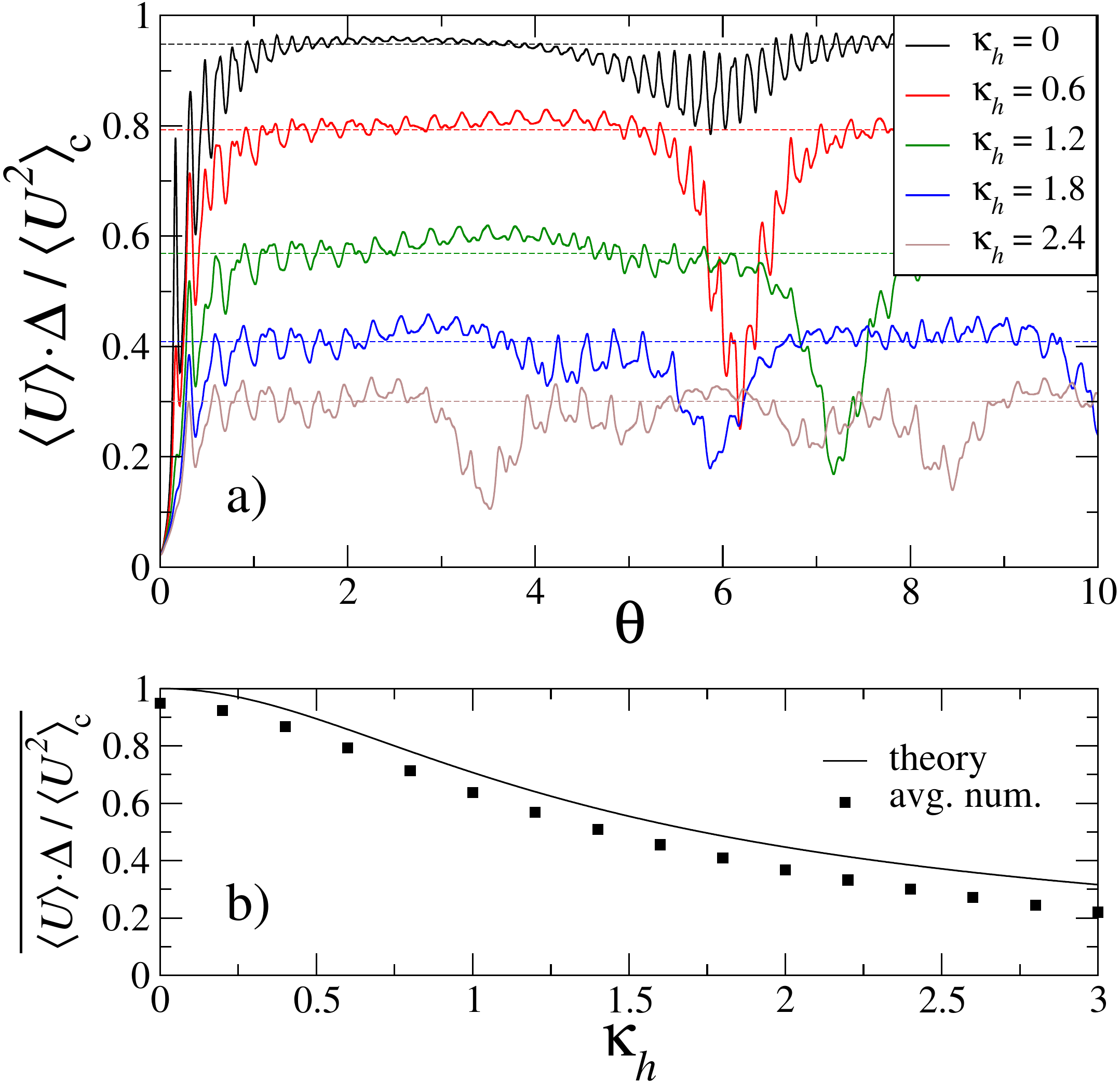}
  \caption{Panel a): ratio between the rescaled average energy of the
    system and its rescaled fluctuations, as a function of $\theta$.
    The various curves are for different values of $\kappa_h$,
    as indicated in the legend (from top to bottom, curves are
    for increasing $\kappa_h$).
    Panel b): comparison between numerical data averaged over the time
    [see horizontal dashed lines in panel a)] and the analytic
    estimate ${\cal U}_1/{\cal U}_2 = 1/\sqrt{1+\kappa_h^2}$, as given
    by Eq.~\eqref{ersquare}.  Data are for an Ising-chain system with
    $L=12$ sites, while all the other parameters are set as in
    Fig.~\ref{Check_HamS_elem}.}
  \label{Check_HamS_ratio}
\end{figure}
%%%%%%%%%%%%%%%%%%%%%%%%%%%%%%%%%%%%%%%%%%%%%%%%%%%%%%%%%%%%%%%%%%%

The discrepancies highlighted above can be amplified when measuring
fluctuations.  Indeed, in Fig.~\ref{Check_HamS_ratio} we show the
ratio between the rescaled average energy of the system $\langle U
\rangle / \Delta$ and its rescaled fluctuations $\langle U^2 \rangle_c
/ \Delta^2$.  The two-level reduction of system $S$ would predict a
value for such ratio which depends only on $\kappa_h$, since it can be
shown that Eq.~\eqref{ersquare} still holds if $v \neq 0$.
Conversely, as displayed in panel a), the full simulation shows a
nontrivial dependence on $\theta$, as well.  The comparison between
the numerical values averaged over $\theta$ (horizontal dashed lines)
and the analytic estimate $\sim 1/\sqrt{1+\kappa_h^2}$ given by
Eq.~\eqref{ersquare} is provided in panel b), as a function of
$\kappa_h$.  Similarly to what observed in Fig.~\ref{Work_FOQT} b) and
Fig.~\ref{EnergyS_FOQT} b), we highlight the emergence of a
discrepancy between the two approaches, which however cannot be
interpreted as a simple offset independent of the value of $\kappa_h$.

\end{document}